\documentstyle[12pt,aaspp4,flushrt,epsf]{article}

\def\be{\begin{equation}}
\def\ee{\end{equation}}
\def\beq{\begin{eqnarray}}
\def\eeq{\end{eqnarray}}
\def\half{{\textstyle{1\over2}}}
\def    \simlt  {\lower.5ex\hbox{$\; \buildrel < \over \sim \;$}}
\def    \simgt  {\lower.5ex\hbox{$\; \buildrel > \over \sim \;$}}

\begin{document}

\title{The Keplerian map for the restricted three-body problem as a model of
comet evolution}

\author{Leonid~Malyshkin$^{1,2}$ and Scott~Tremaine$^{1,3}$}
\affil{	$^1$Princeton University Observatory, Princeton, NJ~08544 \\
	$^2$leonmal@astro.princeton.edu, $^3$tremaine@astro.princeton.edu}
\date{}

\begin{abstract}
We examine the evolution of highly eccentric, planet-crossing orbits in the
restricted three-body problem (Sun, planet, comet). We construct a simple
Keplerian map in which the comet energy changes instantaneously at perihelion,
by an amount depending only on the azimuthal angle between the planet and the
comet at the time of perihelion passage. This approximate but very fast
mapping allow us to explore the evolution of large ensembles of long-period
comets. We compare our results on comet evolution with those given by the 
diffusion approximation and by direct integration of comet orbits. We find
that at long times the number of surviving comets is determined by resonance
sticking rather than a random walk. 

\end{abstract}


\section{Introduction}

The orbits of small bodies in the outer solar system, such as planetesimals or
comets, evolve through gravitational interactions with the giant planets. This
process determines the structure of the Oort Cloud, the rate of depletion of
the Kuiper belt, the total population and spatial distribution of the
Centaurs, and the dynamical properties of observed comets. Understanding the
nature of planet-induced orbital evolution is therefore important for a wide
range of solar system problems.

A particularly important issue is the evolution of highly eccentric orbits
whose perihelia penetrate the region of the outer planets (e.g. long-period
comets). In this case the evolution can be treated approximately as a
one-dimensional random walk in energy; the energy change per perihelion
passage is assumed to be a Gaussian random variable and the rms energy change
is assumed to be small compared to the typical orbital energy. This
``diffusion approximation'' was introduced by Oort (1950) and has been
employed by many authors, including Whipple (1962), Weissman (1978), Yabushita
(1980), Bailey (1984), and Duncan et al. (1987). A closely related approach is
to treat the evolution as a random walk; thus the energy change at perihelion
is assumed to be a random variable but it is not assumed to be small compared
to the orbital energy (Kendall 1961, Arnold 1965, \"Opik 1976, Everhart 1977,
1979, Yabushita 1979, Froeschl\'e and Rickman 1980).  The diffusion or
random-walk approximation provides considerable insight into the evolution of
comet orbits, but neglects potentially important effects such as secular
evolution, resonances, etc.

Direct numerical integrations provide the most accurate way to analyze orbital
evolution. The evolution of long-period comets under the gravitational
influence of the giant planets has been examined by Wiegert and Tremaine
(1998), and extensive integrations of short-period comets and Kuiper-belt
objects have been carried out by Holman and Wisdom (1992) and by Duncan and
Levison (1997). Numerical integrations require more work, which can be
prohibitive, and may yield less insight than simpler methods. Thus it remains
worthwhile to investigate approximate models for planet-induced orbital
evolution.

Clearly the random-walk approximation fails for orbits with semimajor axes
comparable to those of the outer planets, since in this case the energy
changes are correlated at successive perihelion passages. A simple approach
that illuminates this transition is to replace the random energy change at
perihelion in the random-walk approximation with an energy change that depends
on the azimuthal phase of the planet at perihelion passage; thus the energy at
successive perihelion passages is determined by a deterministic twist
map of the form 
\beq 
E_{n+1} & = & E_n + F(\psi_n), \nonumber \\ \psi_{n+1} & = &
\psi_n+\Omega_pP(E_{n+1}),
\label{eq:mapp}
\eeq where $E$ is the specific energy, $\Omega_p$ is the planet's angular
speed, $P(E)\propto |E|^{-3/2}$ is the comet's orbital period, $\psi$ is the
azimuthal angle between the planet and the comet perihelion at the time of
perihelion passage, and $F$ represents the energy changes caused by the planet
at perihelion passage. This map was examined by Petrosky (1986) using a
simple---but unrealistic---sinusoidal form for $F$; Petrosky called
(\ref{eq:mapp}) the Keplerian map, a term that we shall also adopt. Sagdeev
and Zaslavsky (1987) derived the map independently, evaluating $F$ in the
simple but uninteresting limit that the perihelion distance was much larger
than the planet's semimajor axis (they also neglected the indirect term in the
gravitational potential from the planet). Chirikov and Vecheslavov (1989) used
the map to examine the long-term evolution of Halley's comet.

The aim of this paper is to provide a fairly rigorous derivation of the
Keplerian map, to discuss its generalizations, to provide better
approximations for the ``kick function'' $F$ than have been available so far,
and to compare the predictions of the map both to numerical integrations and
the diffusion approximation, to clarify the strengths and weaknesses of each
approach. One of our principal results is that the number of comets
that survive for many orbits is much larger than predicted by the diffusion or
random-walk approximation, a result that we ascribe to ``resonance
sticking''. 

Mostly we shall examine the motion of comets in a planetary system containing
the Sun and a single planet of mass $M_p$ on a circular orbit.  Without loss
of generality we may choose the period and the semimajor axis of the planet to
be unity. With this choice of units the planet's mean motion $\Omega_p=2\pi$,
$G(M_\odot+M_p)=4\pi^2$, $P(E)=2^{3/2}\pi^3/(|E|^{3/2})$ is the comet's orbital
period, and the specific energy of the planet $E_p=-2\pi^2$.  We also write
$m_p\equiv M_p/M_\odot$.

\subsection{The diffusion approximation}

\label{sec:diff}

In this approximation the evolution of the comet orbit is treated as a
one-dimensional random walk in energy, the mean-square energy change per
perihelion passage $D$ is assumed to be independent of energy and $D^{1/2}$ is
assumed to be small compared to the typical comet energy $E$. Then the number
of comets bound to the solar system at time $t$ with energy in the range
$[E,E+dE]$ is $n(E,t)dE$, where 
\be 
{\partial n\over \partial t}=\half D{\partial^2\over{\partial E^2}} 
\left[n\over P(E)\right].
\label{eq:diff}
\ee
A more convenient form is 
\be
{\partial n\over \partial t}=m_p^2 D_0{\partial^2\over{\partial E^2}}
\left[{(-E)}^{3/2}n\right], 
\label{eq:diffa}
\ee
where $D_0$ is independent of the planetary mass,
\be
m_p^2 D_0={1\over 2^{5/2}\pi^3}D = {1\over 2^{5/2}\pi^3}\langle(\Delta E)^2
\rangle,
\label{eq:diff_coeff}
\ee 
and the average $\langle\cdot\rangle$ is taken over the phase of the planetary
orbit at a single perihelion passage.

The solution to equation (\ref{eq:diff}) depends on the boundary
conditions. The simplest assumption is that all comets with positive energy
escape, so $n=0$ for $E\ge0$. In this case the Green's function corresponding
to the initial distribution $n(E,t=0)=\delta(E-E_0)$, $E_0<0$, is (Yabushita
1980) 
\be 
n(E,t)=\frac{2|E_0|^{1/2}}{m_p^2 D_0 t |E|}
\exp\left(-\frac{4}{m_p^2 D_0t}
\left[|E_0|^{1/2}+|E|^{1/2}\right]\right) I_2\left(\frac{8}{m_p^2 D_0 t} 
(E E_0)^{1/4}\right), 
\label{eq:diffsol}
\ee 
where $I_2$ is a modified Bessel function. The total number of surviving
comets after time $t$ is given by the depletion function
\be
N(t)=\int_0^\infty n(E,t)dE = \gamma\left[2,\; 4{|E_0|}^{1/2}/(m_p^2 D_0 t)\right],
\label{eq:diff_N}
\ee 
where $\gamma$ is an incomplete gamma function. At large times $N(t)\to
8|E_0|\left/{(m_p^2 D_0 t)}^2\right.$. Note that equations (\ref{eq:diffsol})
and (\ref{eq:diff_N}) are normalized so that the initial number of comets 
$N(t=0)=1$.

A more accurate approximation is that comets are lost if either $E\ge 0$ or
$E\le E_{min}$; the absorbing boundary at $E_{min}$ arises because comets in
tightly bound orbits may collide with the Sun, evaporate, or evolve much more
quickly under the influence of more massive planets. In this case it is
straightforward to solve equation (\ref{eq:diffa}) numerically to find the
depletion function.

\section{Mapping method}

We examine the orbit of a test particle (the comet) moving in the combined
gravitational field of the Sun and a single planet on a circular orbit (the
restricted three-body problem).  We restrict our attention to comets with
small inclinations, since we are mostly interested in scattering of comets in
the protoplanetary disk. Thus the comet's phase-space position is described by
its semimajor axis $a$, eccentricity $e$, argument of perihelion $\varpi$, and
mean anomaly $\ell$. Alternatively we may use canonical elements. The simplest
of these are the coordinate-momentum pairs $(\ell,\; K=[GM_\odot a]^{1/2})$ and
$(\varpi,L)$, where $L=[GM_\odot a(1-e^2)]^{1/2}$ is the specific angular
momentum.  The azimuth of the planet at time $t$ may be written
$\phi_p(t)=\Omega_pt+\phi_0=2\pi t+\phi_0$.

The equations of motion are described by a Hamiltonian 
\be
H(K,L,\ell,\varpi,t)=-{8\pi^4\over K^2}+m_pH_1(K,L,\ell,\varpi,t).
\ee
For our purposes it is more convenient to use mean anomaly instead of time as
the independent variable. Then the equations of motion are still Hamiltonian
if we use $(\varpi,L)$ and $(t,-E)$ as canonical coordinates and
$-K(L,E,\varpi,t,\ell)$ as the Hamiltonian, where $K$ is defined implicitly by
$E=H(K,L,\ell,\varpi,t)$ and $E$ is the energy. In other words
\be
{dL\over d\ell}={\partial K\over\partial \varpi},\quad {dE\over
d\ell}=-{\partial K\over\partial t},\quad {d\varpi\over d\ell}=-{\partial
K\over\partial L},\quad {dt\over d\ell}={\partial K\over\partial E}.
\label{eq:hameq}
\ee

We shall now focus on long-period comets. These spend most of their time at
distances much greater than the planet's semimajor axis, where they travel on
near-Keplerian orbits around the Sun-planet barycenter. Changes in the comet's
energy and other orbital elements are localized near perihelion, where its
interactions with the planet are strongest. Therefore it makes sense to assume
that the encounter with the planet lasts for only a short time near perihelion
($\ell=0$). Furthermore since $m_p\ll1$ the perturbation from the planet is
weak. Thus we may write the negative of the Hamiltonian as 
\be
K(L,E,\varpi,t,\ell)={4\pi^2\over(-2E)^{1/2}}+m_p\kappa(L,E,\varpi,t)
\delta_{2\pi}(\ell) + \hbox{O}(m_p^2), 
\ee 
where $\delta_{2\pi}(x)=\sum_n\delta(x-2\pi n)$ is the periodic Dirac
delta-function and $\kappa=-4\pi^2(-2E)^{-3/2}\int H_1d\ell$. 
Furthermore, $\kappa$ can only depend on the
time through the planet azimuth $\phi_p(t)$ and can only
depend on the azimuthal angles $\phi_p(t)$ and $\varpi$ through the
combination $\psi=\phi_p(t)-\varpi=2\pi t+\phi_0-\varpi$. Thus we may write
\be 
K(L,E,\varpi,t,\ell)={4\pi^2\over(-2E)^{1/2}}+m_p\kappa(L,E,2\pi
t+\phi_0-\varpi)\delta_{2\pi}(\ell).
\label{eq:jaccc}
\ee 
Equations (\ref{eq:hameq}) and (\ref{eq:jaccc}) imply that the Jacobi constant
\be
J\equiv E-2\pi L
\label{eq:jacobia}
\ee
is conserved, a property that we inherit from the restricted
three-body problem. Note also that $\kappa$ is conserved as the trajectory
crosses $\ell=0$.  

To proceed further we shall adopt a simple functional form for
$\kappa(L,E,\psi)$ that (we hope) captures the most important physics of the
interaction.  A natural first approximation is that $\kappa$ is independent of
$E$ and $L$, so that $\kappa=\kappa(\psi)$ (we discuss more general
forms in \S\ref{sec:disc}). The assumption that $\kappa$ is
independent of energy is reasonable for long-period comets, which all have
near-parabolic orbits in the vicinity of the planets. The assumption that
$\kappa$ is independent of angular momentum is less natural since the
interaction of a long-period comet with the planets depends strongly on its
perihelion distance; however, the same assumption is made in the diffusion
approximation (eq. \ref{eq:diffa}), and conservation of the Jacobi constant
(eq. \ref{eq:jacobia}) ensures that variations in $L$ are small so long as the
comets remain long-period (i.e. $|E|\ll 1$).

Hamilton's equations (\ref{eq:hameq}) can now be integrated from $\ell=2\pi
n-0$ to $2\pi(n+1)-0$:
\beq
\varpi & = & \mbox{const},
\nonumber \\
E_{n+1} & = & E_n + m_p f(\psi_n),
\nonumber \\
L_{n+1} & = & L_n+{m_p\over 2\pi} f(\psi_n),
\nonumber \\
t_{n+1} & = & t_n+2^{3/2}\pi^3(-E_{n+1})^{-3/2},
\nonumber \\
\psi_{n+1} & = & \psi_n+2\pi(t_{n+1}-t_n),
\label{eq:mapping}
\eeq
where the ``kick function'' $f(\psi)=-2\pi d\kappa(\psi)/d\psi$ is 
independent of planet mass, and the map we have derived is essentially
the Keplerian map (\ref{eq:mapp}). 

Thus we have arrived at a symplectic mapping in two dimensions ($\psi, E$),
which depends on the planet mass and the kick function. The angular momentum
$L$ is determined by the energy and the Jacobi constant $J$
(eq. \ref{eq:jacobia}). The derivation of the map was motivated by
long-period comets, whose orbital period is much longer than the planet
orbital period ($P(E)/P(E_p)=(2\pi^2/|E|)^{3/2}\gg1$); however, it should also
provide a fair representation of the behaviour of orbits with shorter periods,
$|E|/(2\pi^2)\sim 1$. 

Mappings such as (\ref{eq:mapping}) can have trajectories of two kinds,
regular and stochastic. The energy of regular orbits will vary only over a
limited range, whereas stochastic orbits can wander over the entire stochastic
region, and in particular they can reach escape energy ($E\ge 0$), which
corresponds to loss of a comet from the solar system.  The comet is also lost
if $L\le 0$, which corresponds to collision with the Sun and occurs if $E\le
J$.

\subsection{The kick function}

To determine the kick function $f(\psi)$ we numerically integrated the orbits
of a set of comets on initially parabolic orbits through a single perihelion
passage. The comets shared a common perihelion distance $q$ and were
distributed uniformly random in argument of perihelion and longitude of
ascending node. The inclinations were chosen at random from a narrow Gaussian
distribution with zero mean and dispersion~$0.1$ radians.  Then we calculated
the energy change $\Delta E=E_{final}-E_{initial}$ in the barycentric frame
and averaged $\Delta E$ over all comets with a given value of $\psi$, which is
the azimuthal angle between the comet and planet at the instant of perihelion
passage.  The kick function is the average energy change normalized by the
planet mass, $f(\psi)=\langle\Delta E\rangle/m_p$.  Figure~1(a) shows a
scatter plot of the normalized energy change $\Delta E(\psi)/m_p$ versus
$\psi$ for 30,000 passages with perihelion distance $q=0.5$. The Figure also
shows the corresponding kick function $\langle\Delta E\rangle/m_p$, averaged
over $2000$ comets at each of $600$ values of $\psi$. As we should expect, the kick
function is approximately odd\footnote{Strictly, the energy change is only odd
to first order in $m_p$.}, $f(-\psi)=-f(\psi)$. The sharp changes in the kick
function correspond to close encounters with the planet, which occur at
\be
\psi=\pm\left[\arccos(2q-1)-2^{1/2}(1-q)^{1/2}(2q+1)/3\right].
\label{eq:coll}
\ee
Kick functions for different perihelion distances are shown in Figure~1(b).
(Everhart (1968) made an impressive early effort to obtain an
empirical fit to the distribution of energy changes caused by planetary
perturbations; however, his results are averaged over $\psi$ and thus are not
useful for our purposes.)

\begin{figure}
\vspace{9cm}
\includegraphics{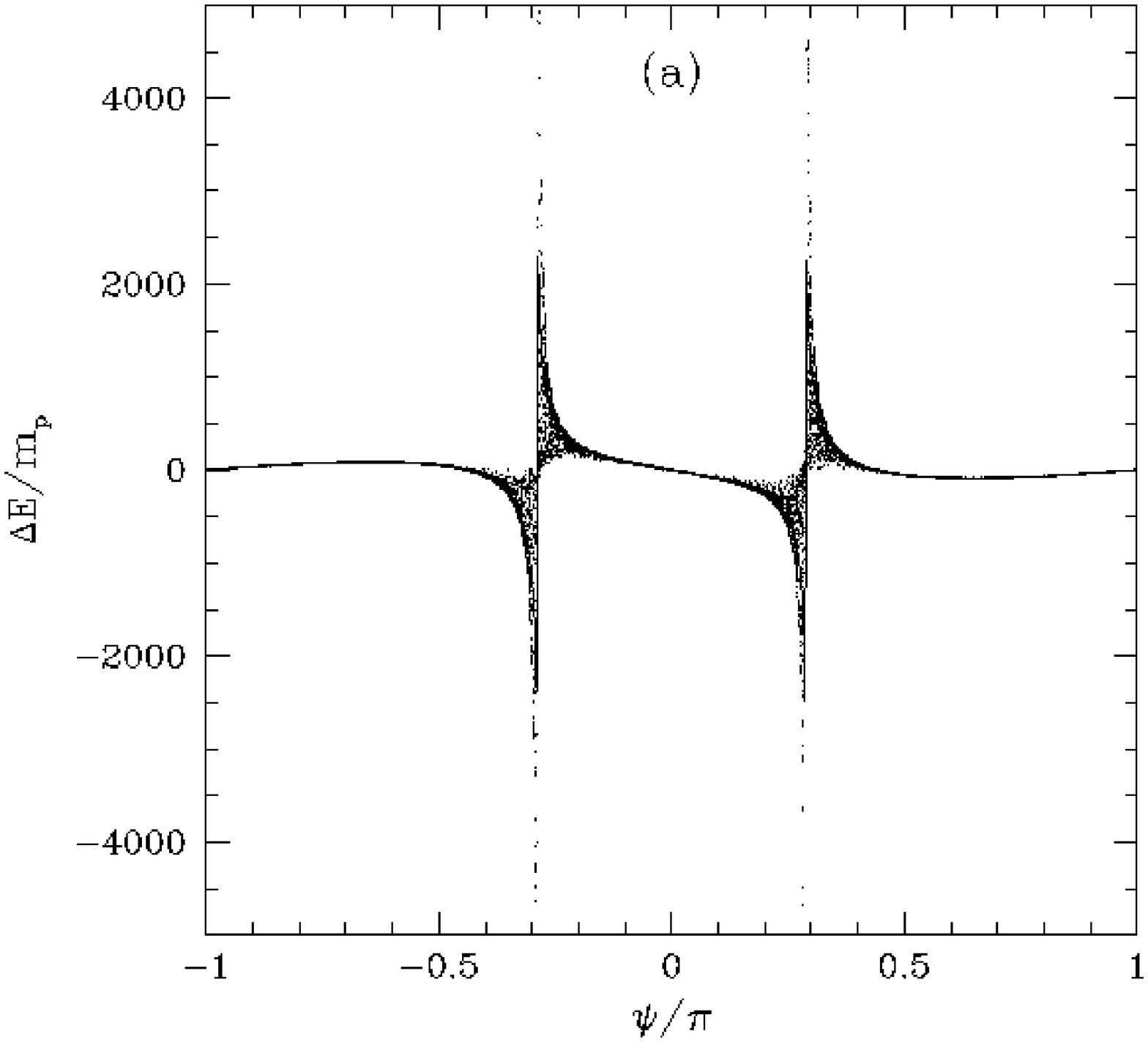}
\includegraphics{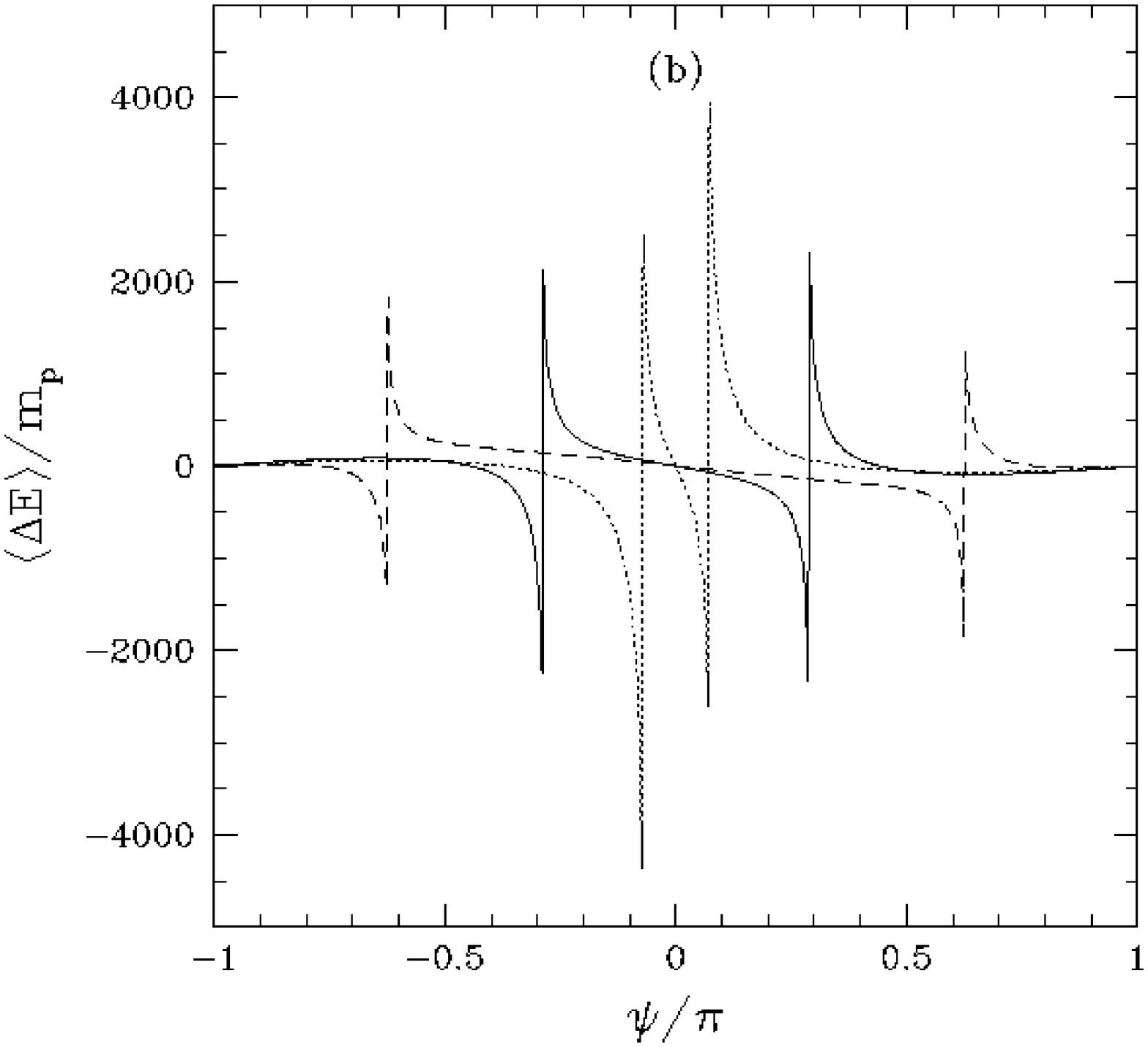}
\caption{
(a) The points show the energy change $\Delta E(\psi)/m_p$ for a sample of
initially parabolic comets with perihelion distance $q=0.5$. The solid line is
the kick function $\langle\Delta E(\psi)\rangle/m_p = f(\psi)$
for this perihelion distance. (b) Kick functions for perihelion distances
$q=0.1$ (dashed line), $q=0.5$ (solid line), and $q=0.9$ (dot line).
It is an artifact of the plotting program that the
curves are not precisely odd in $\psi$ near the spikes.}
\end{figure}

In our mapping (\ref{eq:mapping}) we represent $f(\psi)$
by a continuous interpolation function     
\be
F(\psi)=-F(-\psi)=
\left\{
\begin{array}{ll}
{|\psi-\psi_0|}^{-1/2} \sum\limits_{n=1}^4 C_n^{(-)} 
{(\psi+\pi)}^n & -\pi\le\psi\le\psi_-, \nonumber \\
C_1\psi+C_0 & \psi_-<\psi<\psi_+, \\
{|\psi-\psi_0|}^{-1/2} \sum\limits_{n=1}^4 C_n^{(+)} \psi^n &
\psi_+\le\psi\le 0, \nonumber \\ 
\end{array}
\right .
\ee
where we have divided the interval $[-\pi,0]$ in three parts $[-\pi,\psi_-]$, 
$(\psi_-,\psi_+)$, $[\psi_+,0]$ and define $\psi_-$, $\psi_+$, and $\psi_0$ by 
\be
f(\psi_-)=\min\limits_{\psi\in [-\pi,0]} f(\psi), 
\qquad f(\psi_+)=\max\limits_{\psi\in [-\pi,0]} f(\psi), 
\qquad \psi_0=(\psi_-+\psi_+)/2.
\ee
Of course the function $F(\psi)$ has the same odd symmetry as $f(\psi)$. Note
that only 10 of the 12 coefficients are independent, because of the continuity
constraint at $\psi_-$ and $\psi_+$. 

The coefficients of the interpolation function $F(\psi)$ are given in Table~1,
along with the diffusion coefficient $D_0$ defined by equation
(\ref{eq:diff_coeff}) for initially parabolic orbits with the same inclination
distribution.

\begin{table}
\vspace{1cm}\hspace{10cm}
\begin{center}
\begin{tabular}{||c||c|c|c|c|c||}
\hline
\hline
$q$ & $0.1$ & $0.3$ & $0.5$ & $0.7$ & $0.9$ \\
\hline
\hline
$\psi_-/\pi$ & $-6.26\times 10^{-1}$ & $-4.33\times 10^{-1}$ & $-2.89\times 10^{-1}$ & $-1.73\times 10^{-1}$ & $-7.33\times 10^{-2}$ \\
\hline
$\psi_+/\pi$ & $-6.23\times 10^{-1}$ & $-4.29\times 10^{-1}$ & $-2.86\times 10^{-1}$ & $-1.69\times 10^{-1}$ & $-7.00\times 10^{-2}$ \\
\hline
\hline
$C_1^{(-)}$ & $6.56\times 10^{1}$ & $1.47\times 10^{2}$ & $1.82\times 10^{2}$ & $1.77\times 10^{2}$ & $1.50\times 10^{2}$ \\
\hline
$C_2^{(-)}$ & $7.26$ & $-4.08$ & $-6.22\times 10^{1}$ & $-6.83\times 10^{1}$ & $-5.37\times 10^{1}$ \\
\hline
$C_3^{(-)}$ & $-1.76\times 10^{2}$ & $-9.64\times 10^{1}$ & $-2.42\times 10^{1}$ & $-9.93$ & $-1.07\times 10^{1}$ \\
\hline
$C_4^{(-)}$ & $5.56\times 10^{1}$ & $1.76\times 10^{1}$ & $2.01\times 10^{-1}$ & $-1.15$ & $-2.35\times 10^{-1}$ \\
\hline
\hline
$C_1^{(+)}$ & $-2.40\times 10^{2}$ & $-2.14\times 10^{2}$ & $-3.41\times 10^{2}$ & $-6.48\times 10^{2}$ & $-2.45\times 10^{3}$ \\
\hline
$C_2^{(+)}$ & $-6.84\times 10^{1}$ & $8.83\times 10^{1}$ & $-6.60\times 10^{2}$ & $-2.94\times 10^{3}$ & $-2.83\times 10^{4}$ \\
\hline
$C_3^{(+)}$ & $3.06\times 10^{1}$ & $3.12\times 10^{2}$ & $-1.26\times 10^{3}$ & $-1.08\times 10^{4}$ & $-2.39\times 10^{5}$ \\
\hline
$C_4^{(+)}$ & $1.04\times 10^{1}$ & $1.62\times 10^{2}$ & $-8.11\times 10^{2}$ & $-1.22\times 10^{4}$ & $-6.52\times 10^{5}$ \\
\hline
\hline
$C_0$ & $5.83\times 10^{5}$ & $6.05\times 10^{5}$ & $3.91\times 10^{5}$ & $2.82\times 10^{5}$ & $1.44\times 10^{5}$ \\
\hline
$C_1$ & $2.96\times 10^{5}$ & $4.45\times 10^{5}$ & $4.32\times 10^{5}$ & $5.24\times 10^{5}$ & $6.46\times 10^{5}$ \\
\hline
\hline
$D_0$ & $3.9\times 10^{2}$ & $5.0\times 10^{2}$ & $7.7\times 10^{2}$ & $1.0\times 10^{3}$ & $1.7\times 10^{3}$ \\
\hline
\hline
\end{tabular}
\end{center}
\caption{The coefficients of the interpolation function $F(\psi)$ and 
diffusion coefficients~(eq. \ref{eq:diff_coeff}) for 
different perihelion distances.}
\end{table}

\section{Results}

We examine a map with planet mass $m_p=5.24\times10^{-5}$, which
corresponds to the mass of Neptune, using the kick function for perihelion
distance $q=0.5$ (solid line in Figure~1(b)).  

Figure~2(a) illustrates the behavior of the map. The vertical axis is
$E/2\pi^2$ (the normalization is chosen so that the energy of an orbit with
the planet's semimajor axis is unity), the horizontal axis is $\psi/\pi$, and
each point corresponds to one iteration of the map (one perihelion
passage). The initial energy and azimuth of the orbits plotted are distributed
uniformly random in the range $-2\le E/2\pi^2<0$ and $-\pi\le\psi<\pi$. The
Figure shows the evolution of $100$ comets over a time interval $300$ (recall
that the period of the planet orbit is unity).  Figure~2(c) shows a magnified
view of a smaller energy range, $-1.50\le E/2\pi^2<-1.28$.

Figures~2(b) and 2(d) are the same as Figures~2(a) and~2(c) except that the
evolution of the comet orbits is determined by direct integration of the
restricted three-body problem. The similarity of the corresponding
plots suggests that the map captures most of the relevant dynamics in the
restricted three-body problem. 

The figures exhibit two main classes of orbit: (i) isolated islands consisting
of regular orbits; (ii) a single connected stochastic orbit (the ``stochastic
sea''). The regular orbits are in resonance with the planetary orbital period.
The future evolution of any point in the stochastic sea always terminates with
escape ($E>0$) or collision with the Sun ($L\le0$). The lower limit of the
stochastic sea is at $E=J=E_{init}-2\pi L_{init}$. As the energy approaches
escape the fraction of phase space occupied by the stochastic sea grows.
Notice that there are no KAM surfaces that extend across all values of the
azimuthal angle $\psi$; the reason is that orbits that suffer close encounters
with the planet (eq. \ref{eq:coll}) are always stochastic. Thus stochastic
orbits can escape from any energy range. Many of these features are
reminiscent of the Fermi map with a non-differentiable forcing function
(Lichtenberg and Lieberman 1992).

\begin{figure}
\vspace{16cm}
\includegraphics{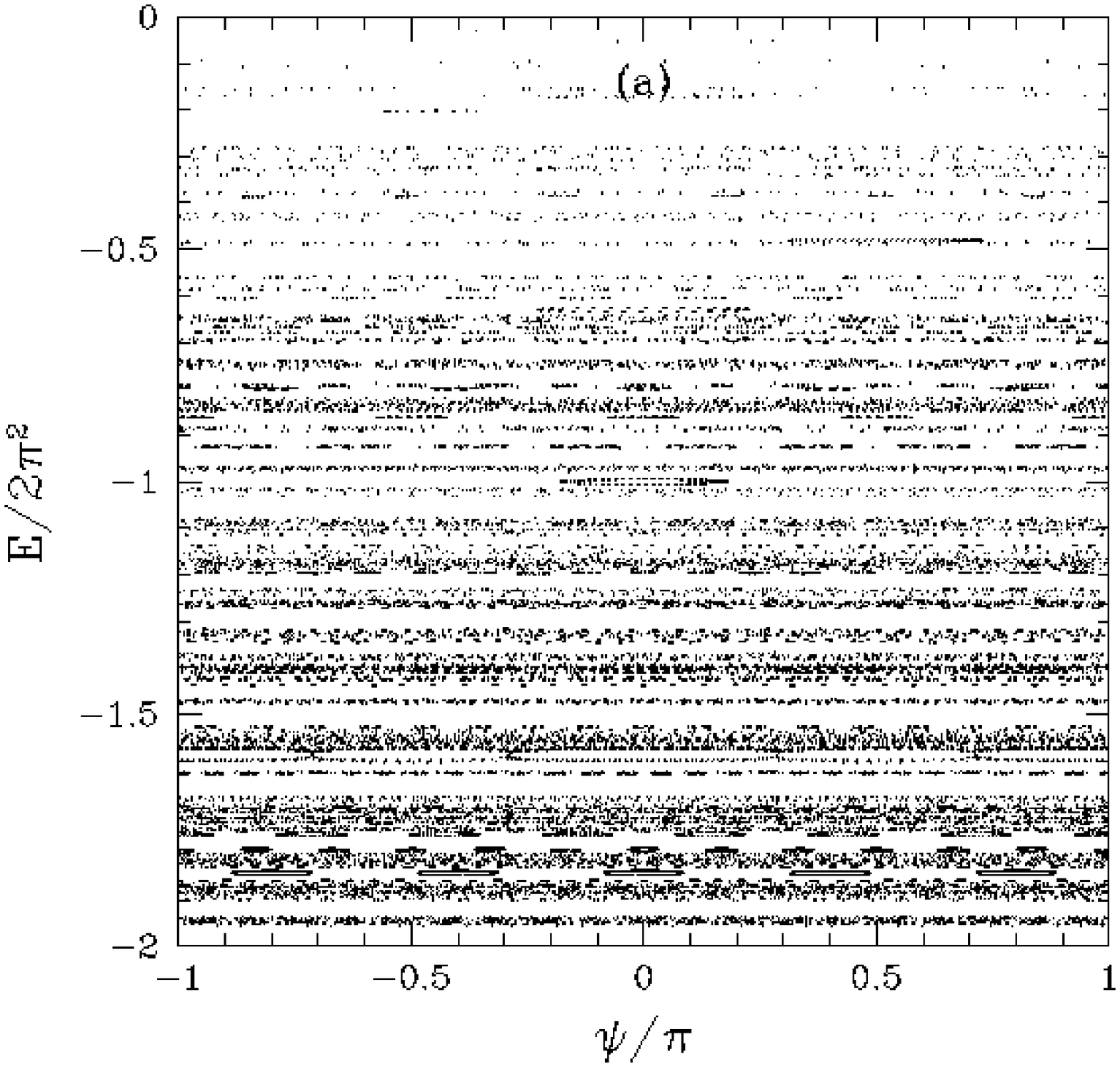}
\includegraphics{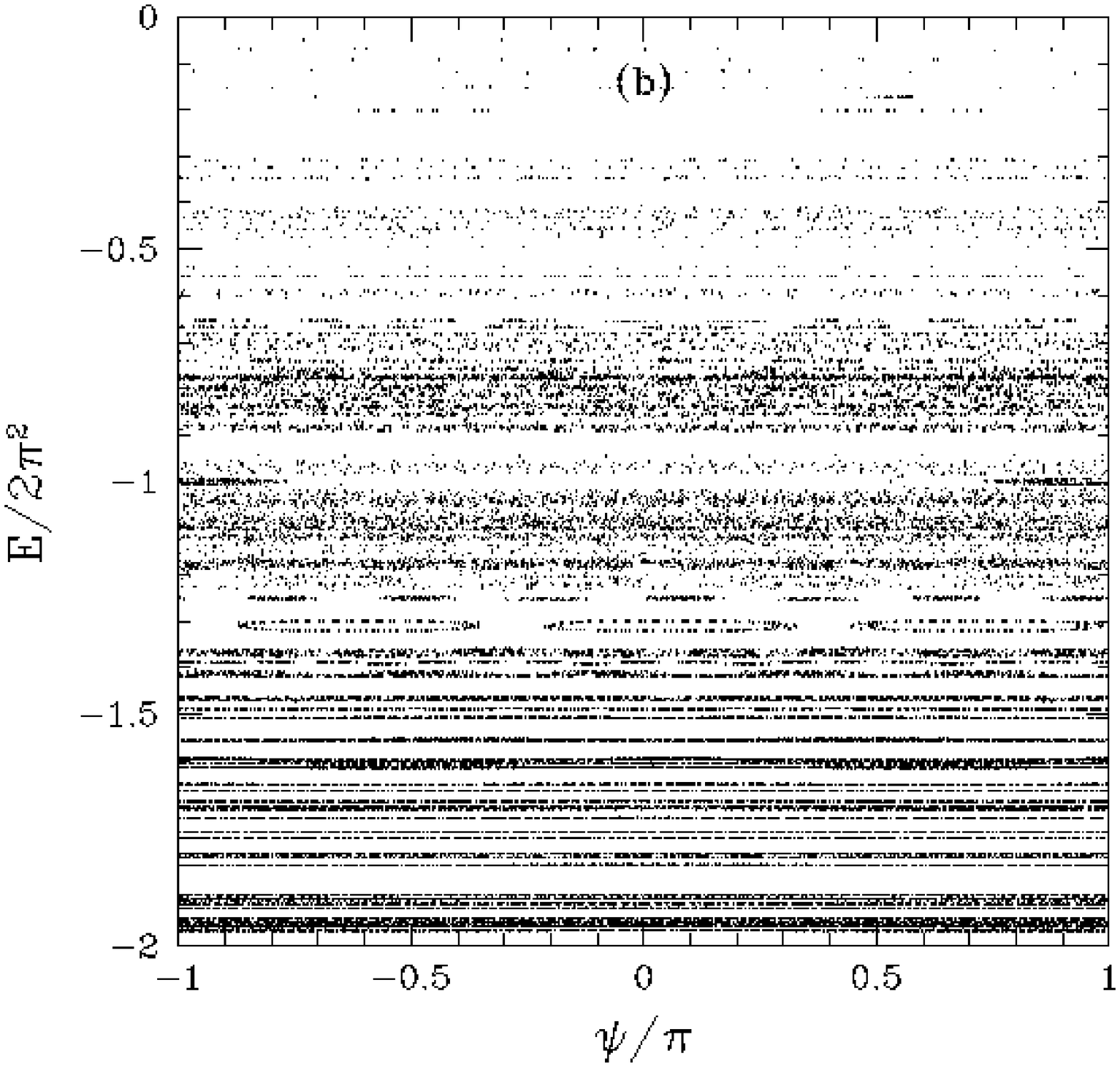}
\includegraphics{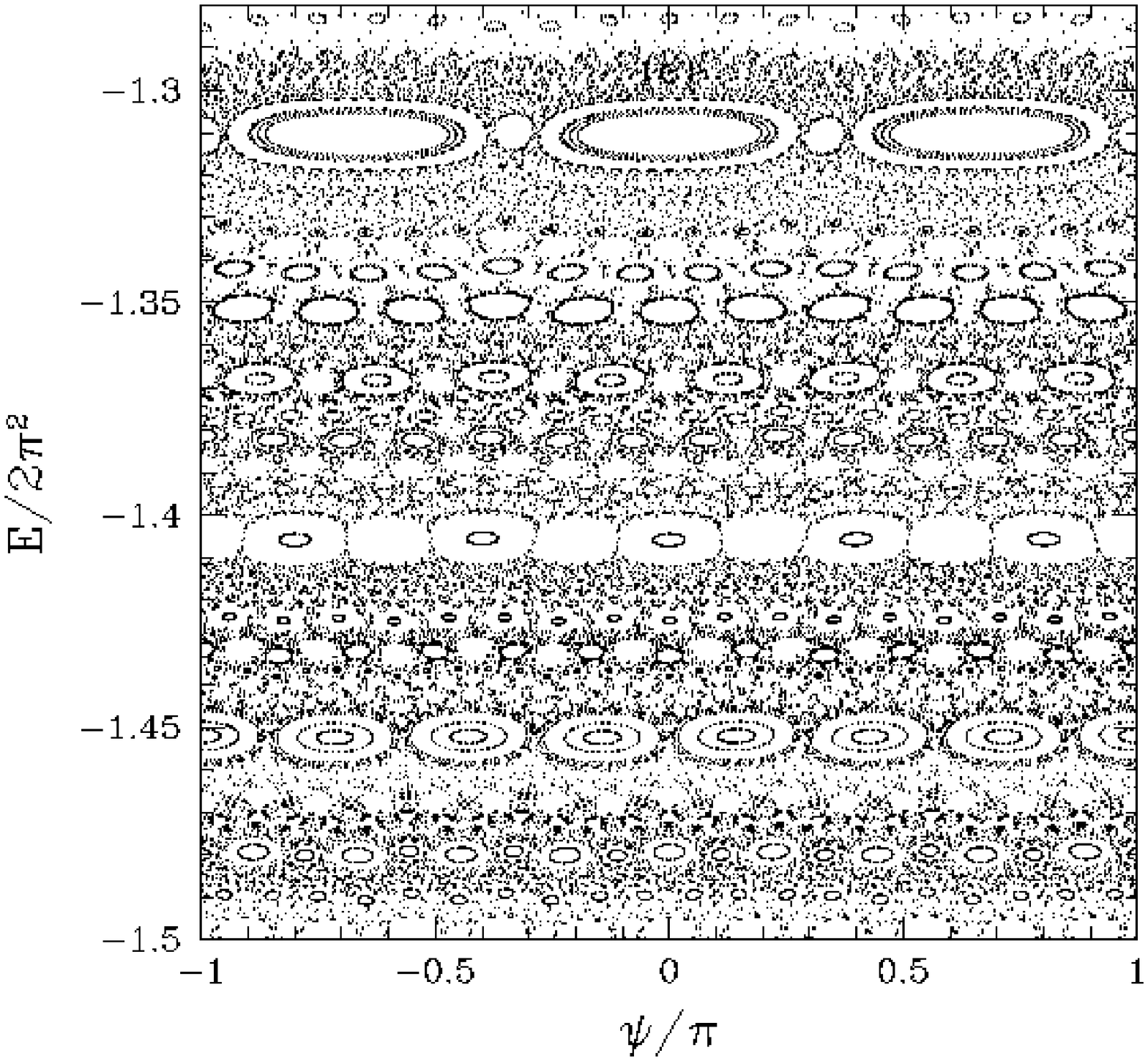}
\includegraphics{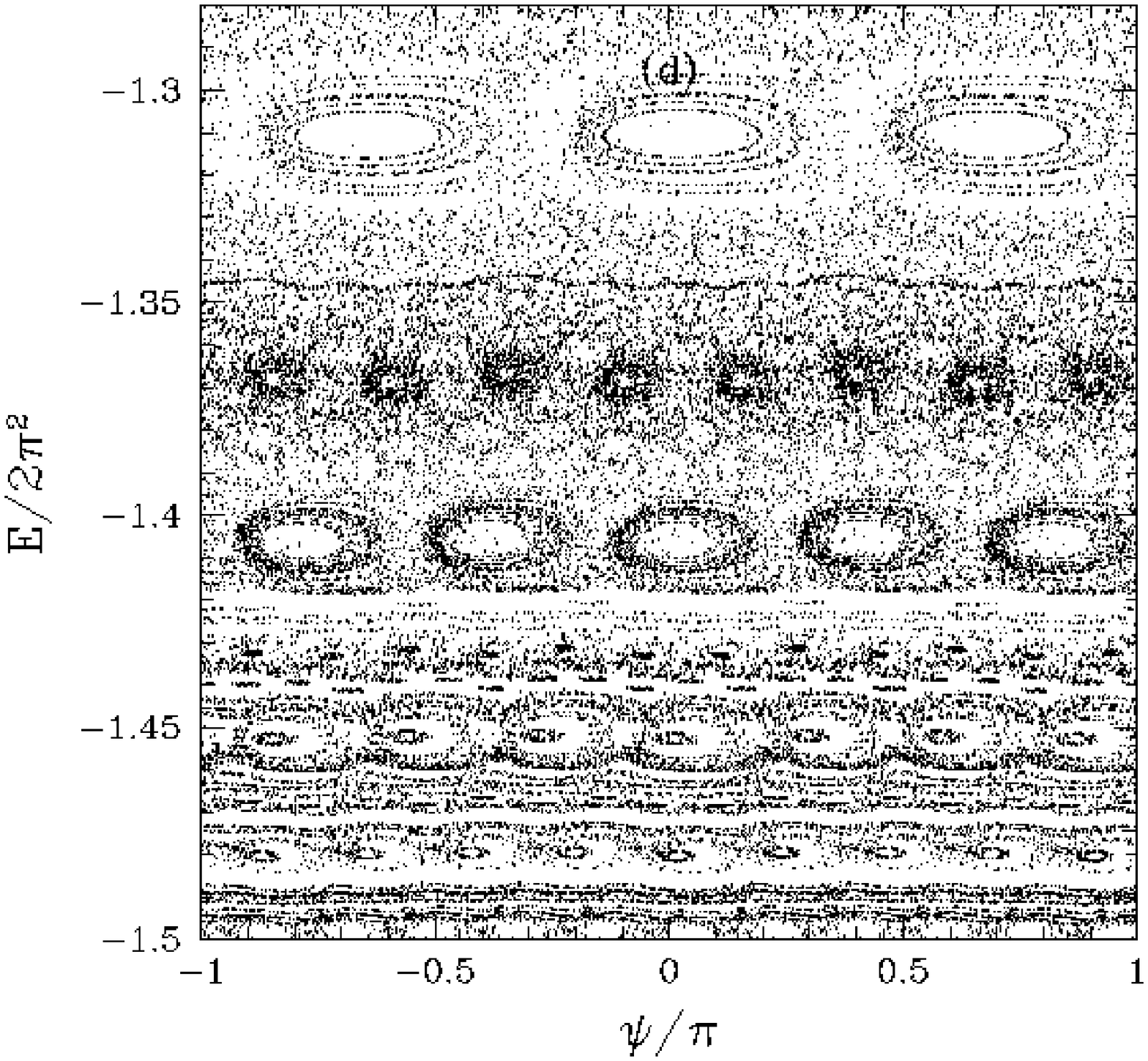}
\caption{ (a) and (c) The perihelion
passages of $100$ comets over $300$ planetary orbital periods (``Neptune
years''), as determined from the map (\ref{eq:mapping}) with perihelion
distance $q=0.5$.  The comets are initially uniformly distributed in energy
and azimuth and the planet mass is $m_p=5.24\times10^{-5}$. (b) and
(d) The same as figures (a) and (c), computed by direct numerical integration
of the comet orbits.}
\end{figure}

To explore the evolution of planet-crossing comets over much longer times we
have used the map to follow $150,000$ comets with initial energy
$E/2\pi^2=-0.242$ (corresponding to semimajor axis $4.13$ in units of the
planetary semimajor axis) and initial azimuth $\psi$ distributed uniformly
random in $[0,2\pi)$. The initial semimajor axis was chosen so that all of the
comets were in the stochastic sea and thus can escape. The comets are lost if
they reach $E=0$ (escape) or $E/2\pi^2=-2.181$ (collision with the Sun, as
determined from (\ref{eq:jacobia})). We followed the comets for $2.7\times
10^7$ time units, which corresponds to the age of the solar system, 4.5 Gyr,
if the planet has Neptune's orbital period. After this time $3.7\%$ of the
comets survived.  Of the comets that are lost, $96.9\%$ escape and the
remainder collide with the Sun. To show the phase portrait of the survivors,
we plot every tenth perihelion passage for a time interval of $1000$ Neptune
years (Figure 3a). To magnify the detail, we also plot all of the perihelion
passages in the interval $-1.50\le E/2\pi^2\le -1.28$ (Figure 3b).

\begin{figure}
\vspace{9cm}
\includegraphics{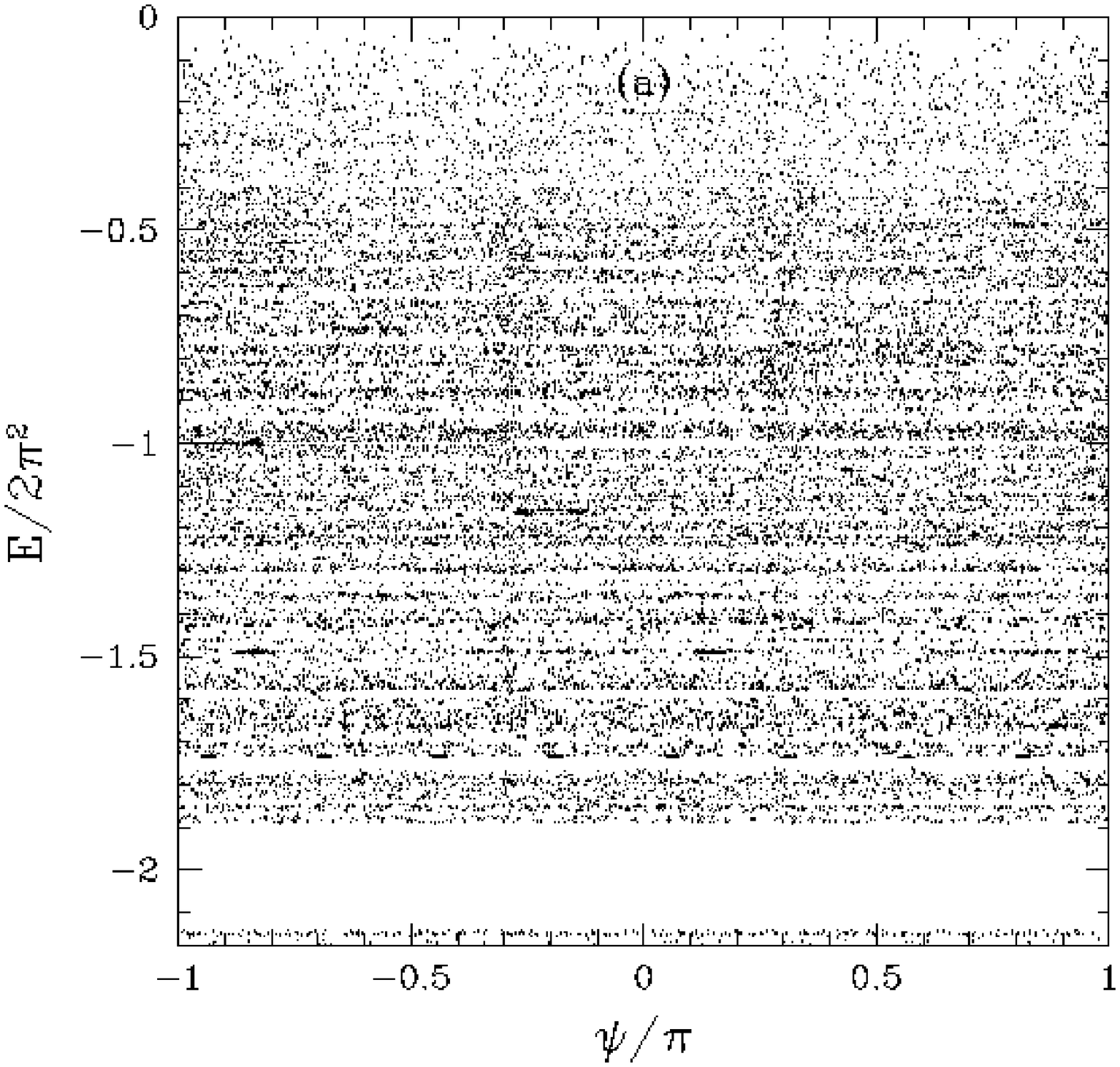}
\includegraphics{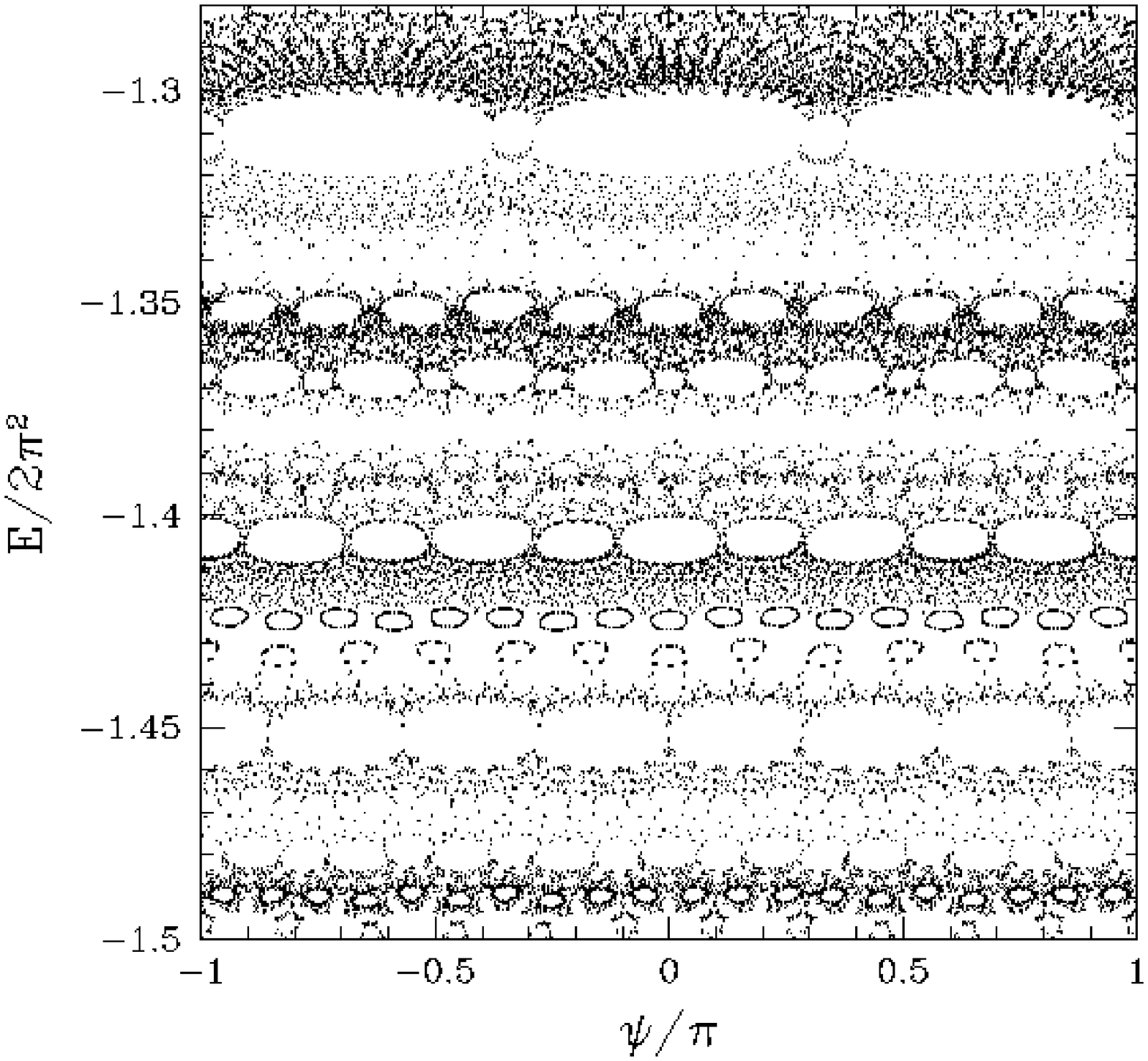}
\caption{
The final phase portrait of $150,000$ comets, after 4.5 Gyr. The planet has
Neptune's mass and orbital period, and the comets initially lie in the
stochastic sea, at $E/2\pi^2=-0.242$ (semimajor axis $4.13$ in Neptune units).
(a) Every tenth perihelion passage of all surviving comets, over an interval
of 1000 Neptune years. The mapping structure for $E/2\pi^2\simlt
-1.8$ is unreliable since very few comets penetrate to this region (for
example, only 3\% of the comets are lost through collision with the Sun at
$E/2\pi^2=-2.181$; all the rest are lost by escape to $E\ge 0$).  
(b) A magnified view showing all perihelion passages in the interval 
$-1.50 \le E/2\pi^2 \le -1.28$.}
\end{figure}

It is worthwhile to compare Figures 2(c) and 3(b). The most striking
difference is that the islands of regular orbits are empty in 3(b); this of
course is because the comets that are shown in 3(b) had initial conditions in
the stochastic sea, whereas those in 2(c) were chosen randomly. A more
interesting difference is that many of the points in Figure 3(b) are
concentrated near the shores of the resonant islands (this is especially
noticeable near $E/2\pi^2\simeq -1.30$, $-1.41$, $-1.43$ and $-1.49$). This is
the well-known phenomenon of resonance sticking: stochastic orbits can be
trapped for extended periods in the forest of tiny resonant islands near the
shore of the stochastic sea (e.g. Karney 1983, Meiss 1992). We explore this
phenomenon further at the end of this section.

To examine the energy distribution of the surviving comets we divided the
sample of 150,000 comets into two equal parts. In Figure 4(a) and 4(b) we plot
the energy distribution of the surviving comets from each sub-sample after 4.5
Gyr; in particular we plot the normalized distribution of the number of
perihelion passages per unit energy interval per unit time, i.e. the function
\be 
\mu(E,t)=\frac{n(E,t)/P(E)}{\int\limits^\infty_0 n(E,t)/P(E)\, dE}
\label{eq:energy_distribution}
\ee
(see definitions in \S \ref{sec:diff}). We see that the curves in the two
panels have different shapes, which shows that most or all of the fine
structure is statistical noise.

\begin{figure}
\vspace{12.5cm}
\includegraphics{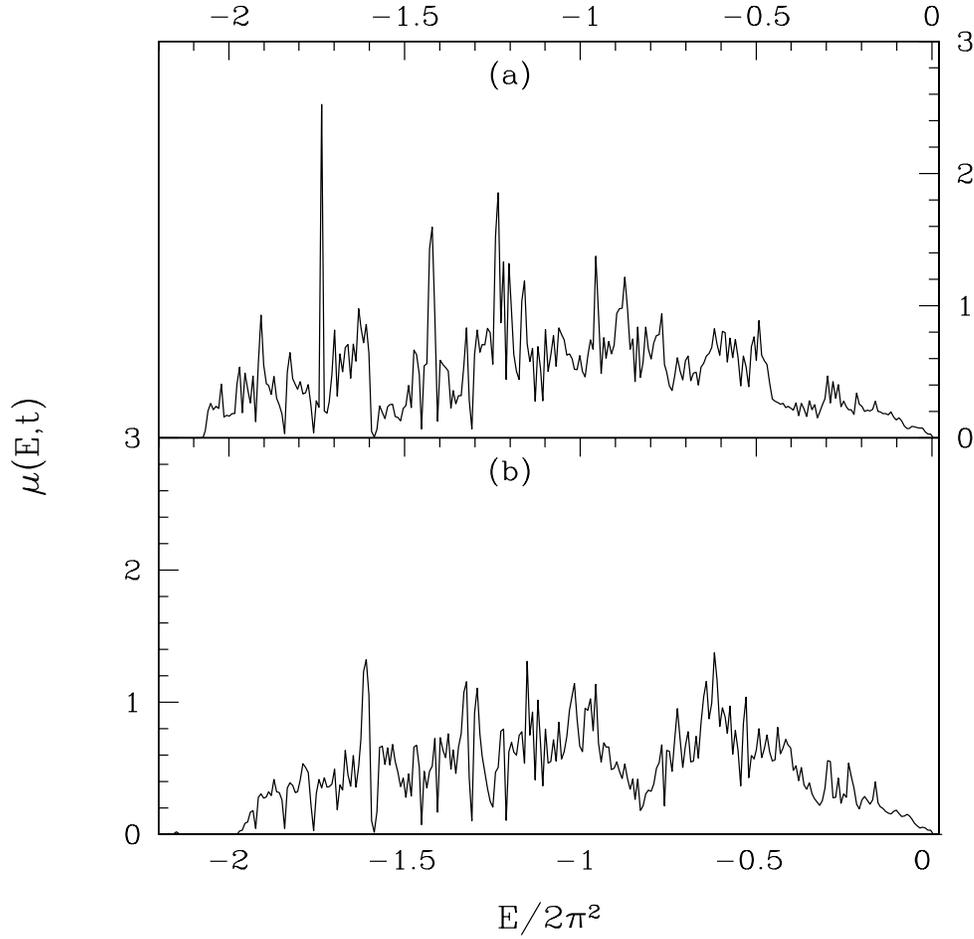}
\caption{ The energy distribution of the surviving comets from Figure 3,
divided into two equal sub-samples. The energy distribution has been averaged
over $10^5$ Neptune years to smooth out short-term fluctuations. The spikes
are generally not common to the two plots and hence represent statistical
noise.}
\end{figure}

Figures 5(a) and~(b) show the depletion function $N(t)$ as a function of time
for several different initial energies. Each curve is based on an initial
sample of $20,000$ comets. The survival fraction at large times in these plots
is unrealistic, as it consists almost entirely of comets that have been kicked
into near-escape orbits of very long period (in practice these comets would be
removed by tidal forces). The expected asymptotic behavior as $t\to\infty$ is
easy to derive heuristically. Let us imagine giving a set of bound comets a
broad distribution of energy kicks at $t=0$. Since the distribution of kicks
is smooth, the number density of comets as a function of energy is flat near
$E=0$, $n(E)dE=\hbox{const}\times dE$. The distribution of orbital periods is
then $n(P)dP=n(E)|dE/dP|dP=\hbox{const}\times P^{-5/3}dP$.  These comets 
remain bound until their second perihelion passage, after which they will
normally be ejected within a relatively short time. Thus the expected
asymptotic shape is
\be N(t)\propto \int\limits_t^{\infty} n(P)\, dP\propto t^{-2/3}.
\label{eq:asymp}
\ee 
This power law is shown by an arrow on Figure~5(b) and accurately
reproduces the behavior of the map.

\begin{figure}
\vspace{9cm}
\includegraphics{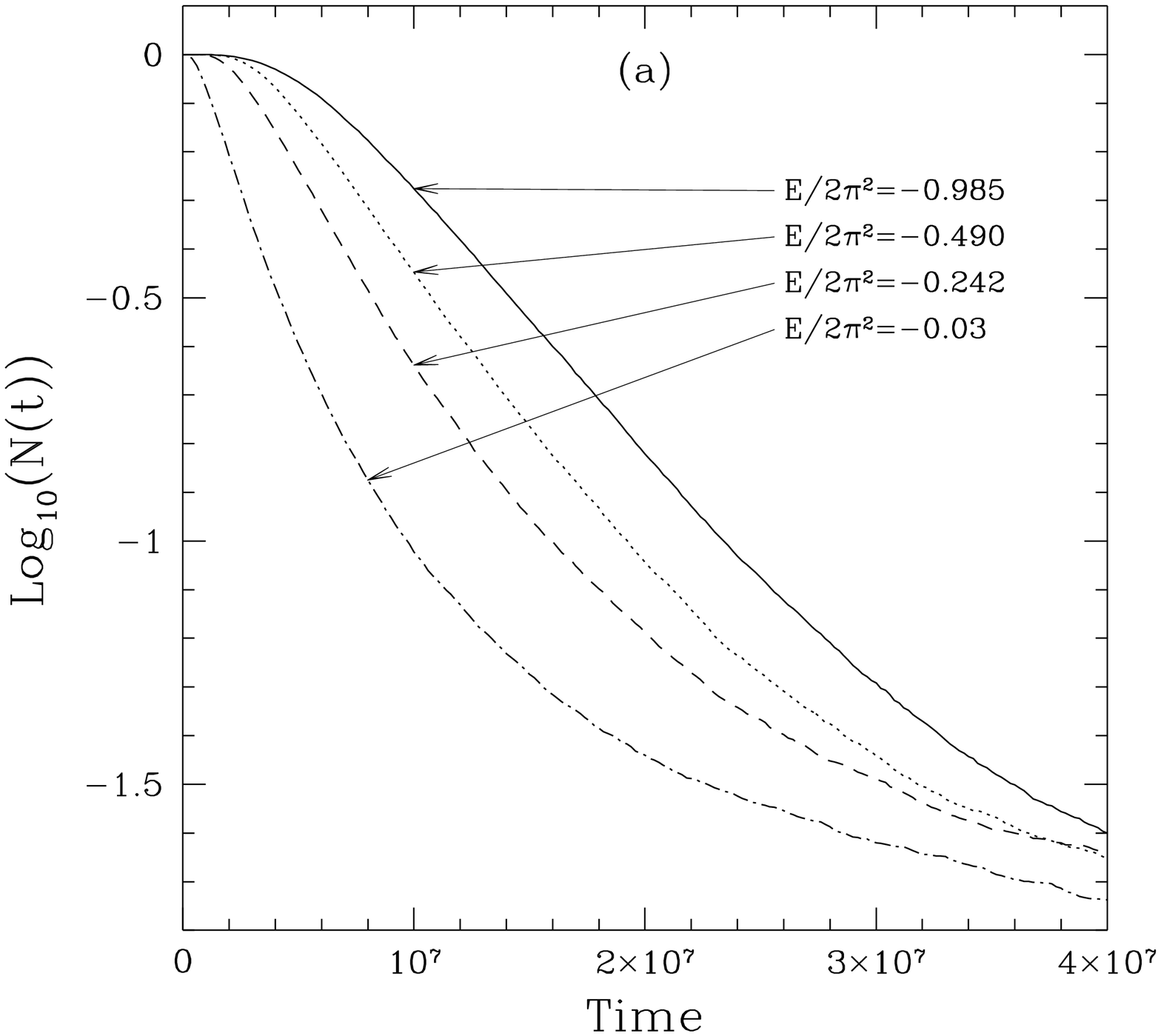}
\includegraphics{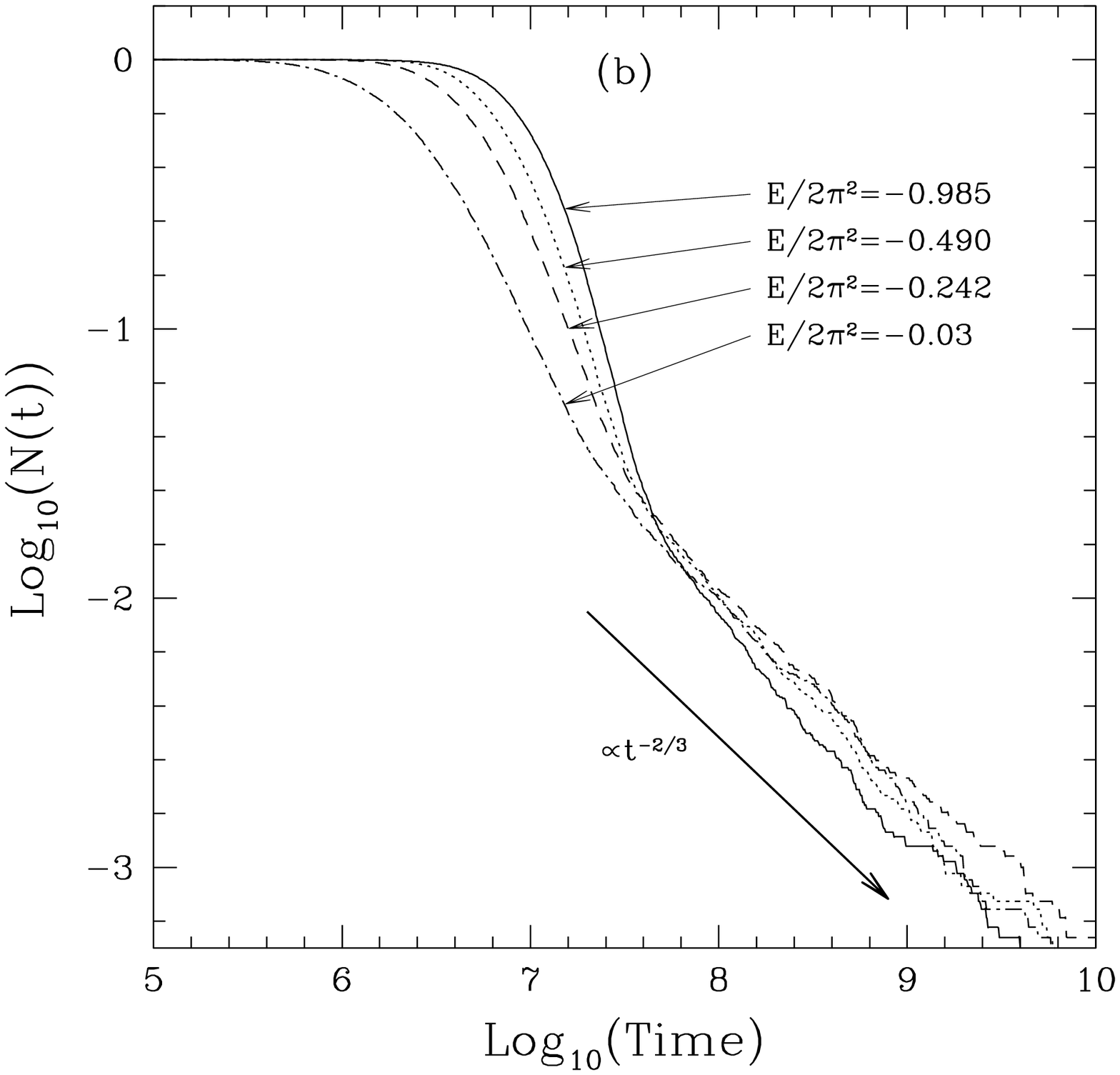}
\caption{ The logarithm of the fraction of surviving comets in the map
(\ref{eq:mapping}) as a function of (a) time and (b) log of the time. Each
sample initially contained $20,000$ comets uniformly distributed in azimuth,
with initial energies $E/2\pi^2=-0.985$, $E/2\pi^2=-0.490$, $E/2\pi^2=-0.242$
and $E/2\pi^2=-0.03$. The initial energies are chosen so that all comets lie
in the stochastic sea. The slope of the asymptotic law~(\ref{eq:asymp}) is
shown by an arrow.}
\end{figure}

A more realistic map can be obtained by choosing a slightly negative escape
energy, to crudely represent the effects of Galactic tidal forces
(e.g. Heisler and Tremaine 1986). In Figures 6(a) and (b) we plot $N(t)$ as
determined by mapping 500,000 comets with initial energy $E/2\pi^2=-0.242$ and
random initial phases. The two solid lines correspond to two different choices
for the escape energy, $E_{esc}/2\pi^2=0$ and $-0.03$; the latter corresponds
to the more realistic case where a comet escapes from the influence of the
giant planets when it reaches a semimajor axis $a\sim 1000\mbox{AU}$ where
galactic tides and passing stars begin to have significant effects (Duncan,
Quinn and Tremaine 1987).  When the escape energy is non-zero $N(t)$ decays
approximately exponentially [Figure 6(a)] until $t_s\sim 5\times10^7$ and
thereafter as $t^{-k}$, $k\approx 1.3$ [Figure 6[b]). We interpret this
late-time behaviour as the result of resonance sticking.

We may also compare the depletion function in the map to the predictions of
the diffusion approximation (eq. \ref{eq:diffa}).  The dotted line shows the
analytic prediction from equation (\ref{eq:diff_N}), using the diffusion
coefficient $D_0$ from Table~1.  As we indicated in \S \ref{sec:diff}, a more
accurate approximation is that comets are lost either when $E\ge 0$ (escape)
or $E/2\pi^2<J/2\pi^2=-2.181$ (collision with the Sun). We have also solved
equation (\ref{eq:diffa}) numerically with these boundary conditions to
provide a second prediction of $N(t)$ (dashed line).  For reference, we show
by solid arrows on Figure~6(b) a power-law depletion law $N(t)\propto t^{-k}$
with $k=1.3$, $k=\frac{2}{3}$ as predicted by equation (\ref{eq:asymp}), and
$k=2$ as predicted by the diffusion approximation with $E_{min}\to-\infty$
(eq. \ref{eq:diff_N}).

\begin{figure}
\vspace{9cm}
\includegraphics{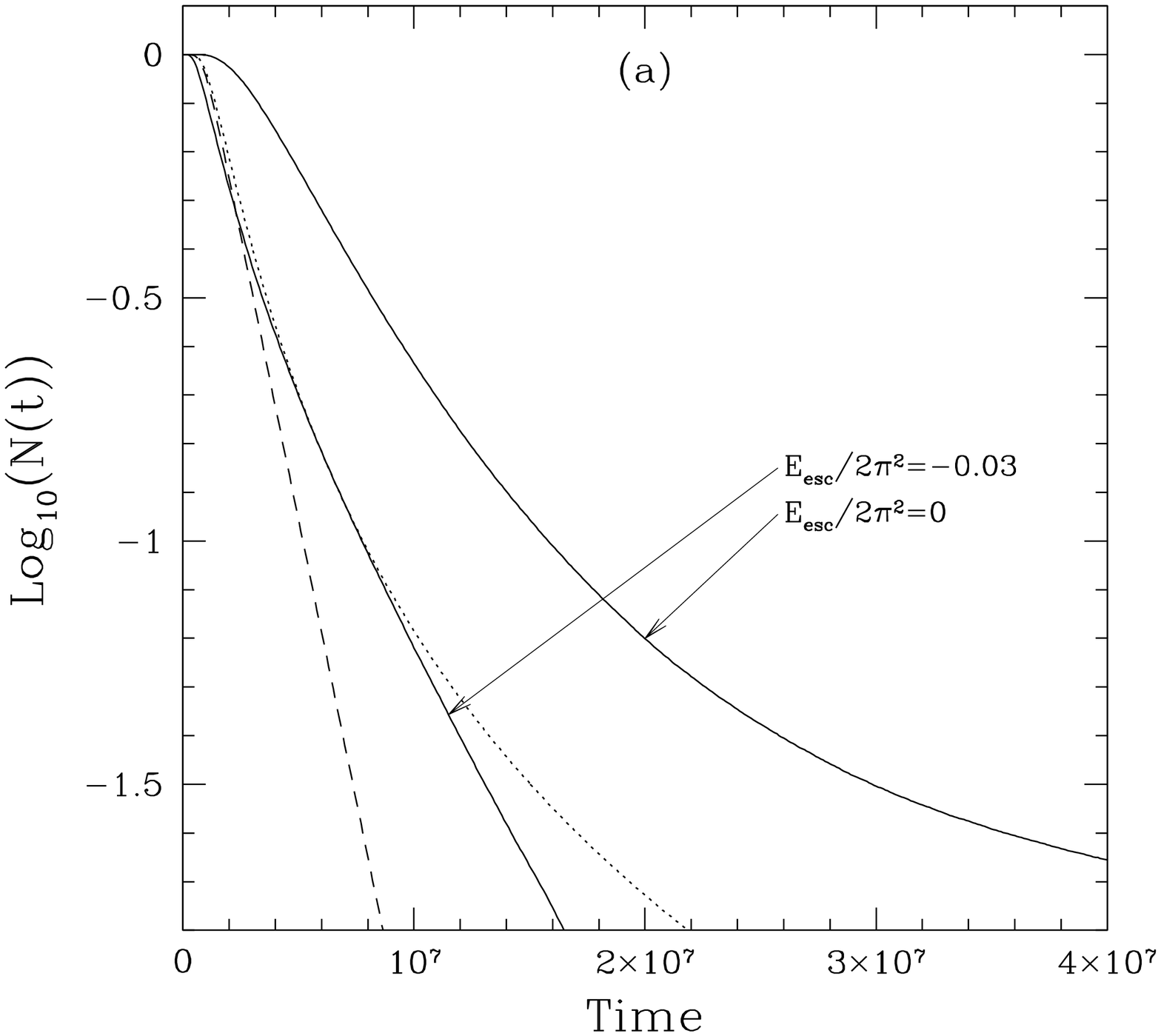}
\includegraphics{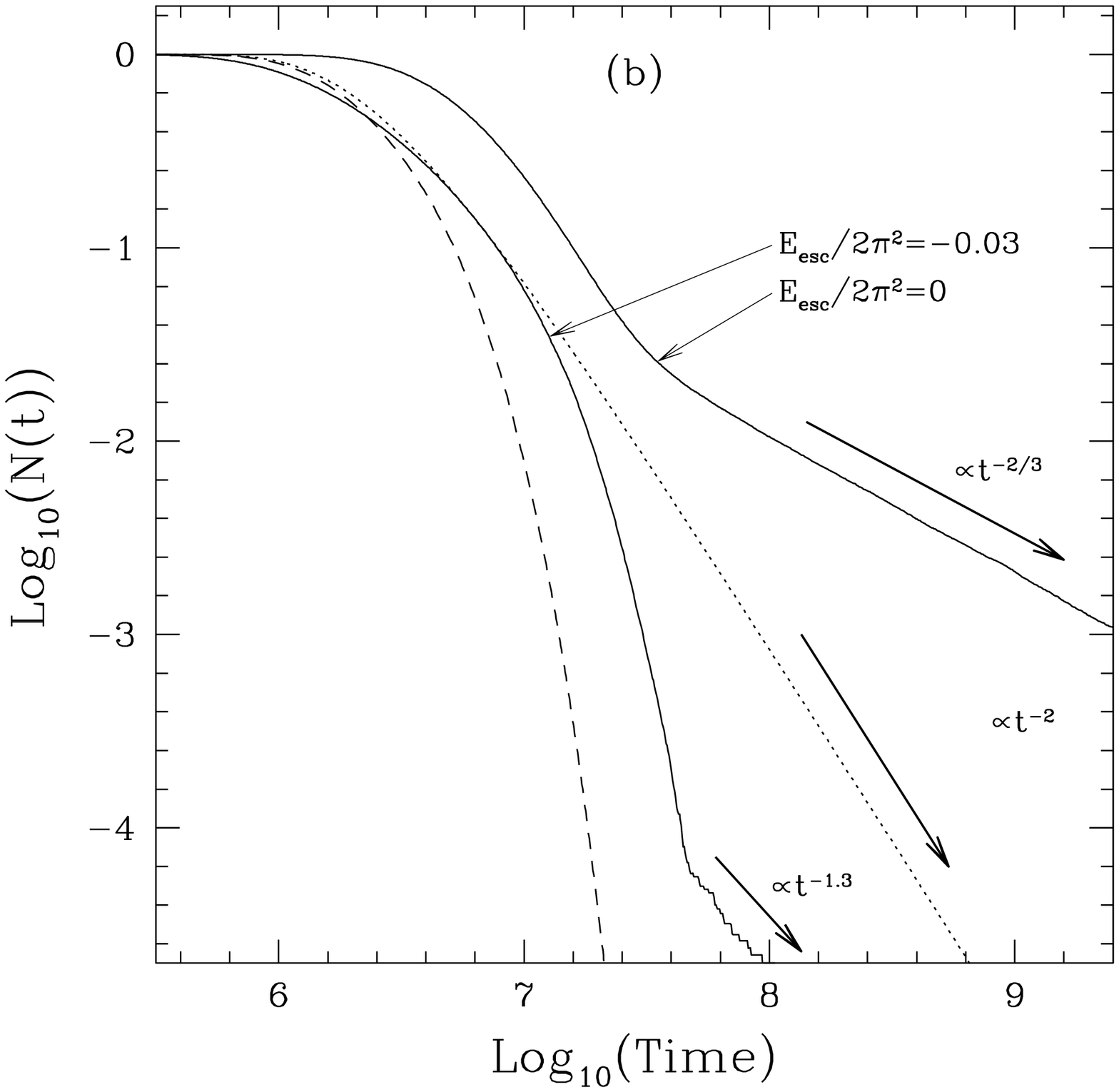}
\caption{The solid lines show the logarithm of the fraction of surviving
comets in the map (\ref{eq:mapping}) as a function of time (a) and logarithm
of time (b). The sample initially contained 500,000 comets with energy
$E/2\pi^2=-0.242$; these were followed for two values of the escape energy,
$E_{esc}/2\pi^2=0$ and $E_{esc}/2\pi^2=-0.03$. We also plotted two predictions
from the diffusion approximation of \S \ref{sec:diff}. The dotted line is for
a single absorbing boundary at $E=0$ (escape) and the dashed line is for
two absorbing boundaries at $E=0$ (escape) and $E/2\pi^2=J/2\pi^2=-2.181$. The
asymptotic falloffs $N(t)\propto t^{-2/3}$, $N(t)\propto t^{-2}$ and
$N(t)\propto t^{-1.3}$ are shown by arrows.}
\end{figure}

We close this section by exploring the phenomenon of resonance sticking in
more detail. To do so, we map $5\times 10^6$ comets with initial energy
$E/2\pi^2=-0.49$, and consider that a comet is lost anytime it reaches the
nearby boundaries $E_{min}/2\pi^2=-0.59$, $E_{max}/2\pi^2=-0.39$ (focusing on
a smaller energy interval provides a more sensitive probe of sticking, since
the diffusion time is reduced for fixed diffusion coefficient). The resulting
depletion functions for the mapping~(eq. \ref{eq:mapping}) and the diffusion
approximation~(eq. \ref{eq:diffa}) are shown in Figure~7 by solid and dashed
lines respectively. We see that for $t\simlt 7\times 10^5$ the depletion
for the mapping is exponential, just as predicted by the diffusion
approximation, although the rate of depletion in the map is slower than the
diffusive prediction. However, for $t\simgt t_s\sim 10^6$ the
map again shows a power-law depletion, $N(t)\propto t^{-1.3}$, instead of the
exponential depletion predicted by the diffusion approximation.  Figure~8(a)
shows the map over 1000 Neptune years for the $67$ comets that survived for
$t=5\times 10^6$. Most of these comets are at the boundary of resonant islands,
confirming that the power-law tail of the depletion function is due to
resonance sticking.

Indirect evidence for resonance sticking is also seen in the energy correlation
function $p(\Delta E,\Delta t)$, which gives the probability that a comet
suffers an energy change $\Delta E$ over a time interval $\Delta t$. For
$|\Delta E/E|\ll 1$ the diffusion approximation predicts 
\be
p(\Delta E,\Delta t)=
\frac{1}{\sqrt{2\pi}\,\sigma}\exp{\left(-{\Delta E}^2/2\sigma^2\right)},
\label{eq:gaussian}
\ee 
\be
\sigma^2=\frac{D\Delta t}{P(E)},
\label{eq:sigma}
\ee 
where $P(E)$ is the comet's orbital period and $D$ is the mean-square energy
change per perihelion passage~(eq. \ref{eq:diff_coeff}).  We have plotted
equation~(\ref{eq:gaussian}) as a dotted line on Figure 8(b).  We have also
plotted the correlation function for the map, at time intervals $\Delta t=20$,
60 and 100 (to estimate the correlation function we followed $10^6$ comets
with initial energy $E/2\pi^2=-0.49$ for $10^6$ Neptune years; we
calculated the normalized distribution of energy changes after every $\Delta
t$ time units and averaged these distributions to obtain the correlation
function).  We see that as $\Delta t$ increases, the correlation function for
the map becomes more and more sharply peaked, which is a sign of resonance
sticking.  A possible concern is that our results are biased by a selection
effect: we only followed comets so long as they remained inside a narrow
energy band, $-0.59< E/2\pi^2<-0.39$. To check whether this bias is present we
have constructed correlation functions using a Gaussian random walk with
mean-square energy step given by the diffusion coefficient $D$ and the same
boundary conditions; these are shown in Figure 8(b) as dashed curves. In fact
the dashed curves are so similar that they appear as a single near-Gaussian
curve, which confirms that the peak seen in the correlation function for the
map is not a result of the absorbing boundaries. 

\begin{figure}
\vspace{9cm}
\includegraphics{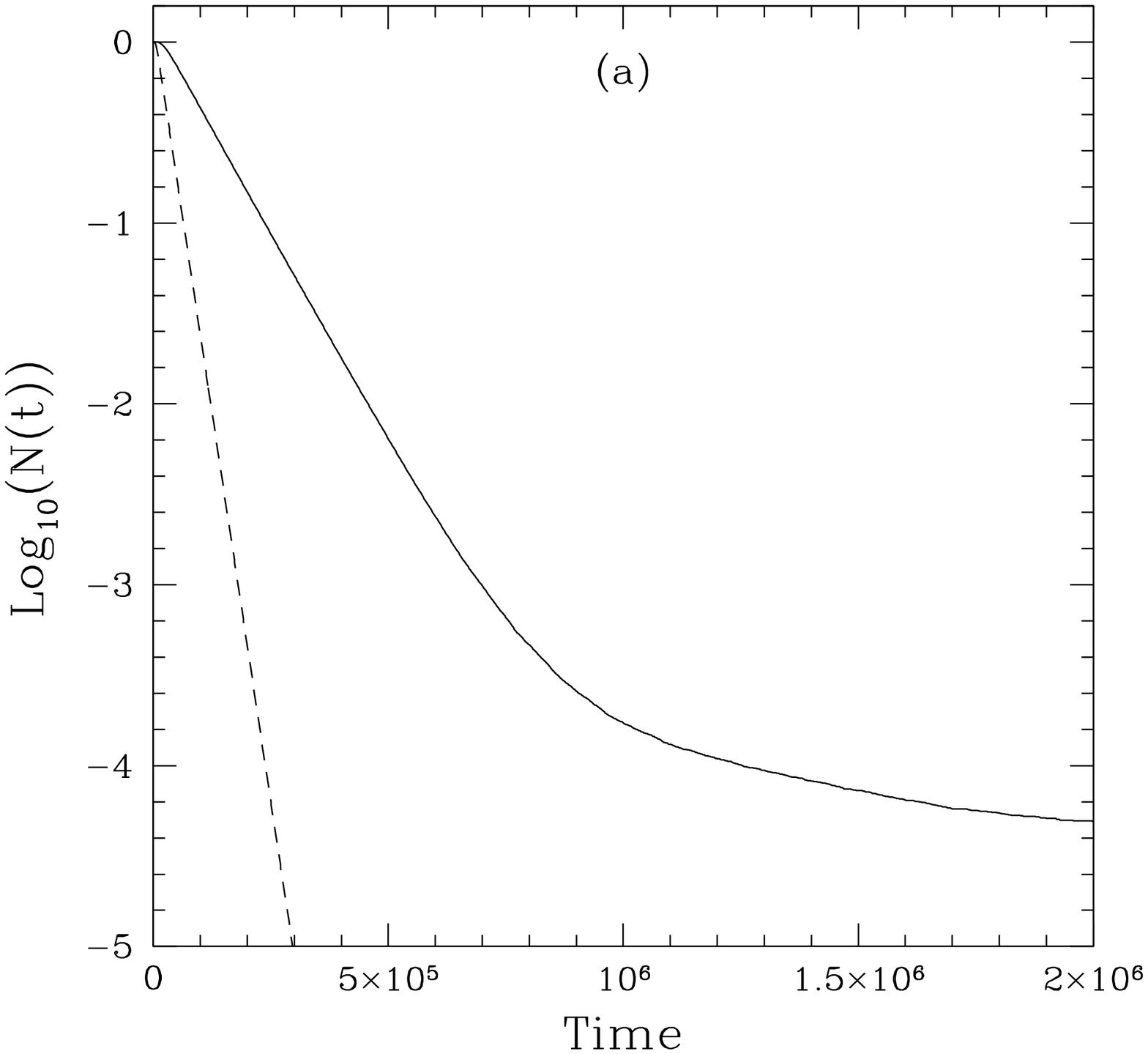}
\includegraphics{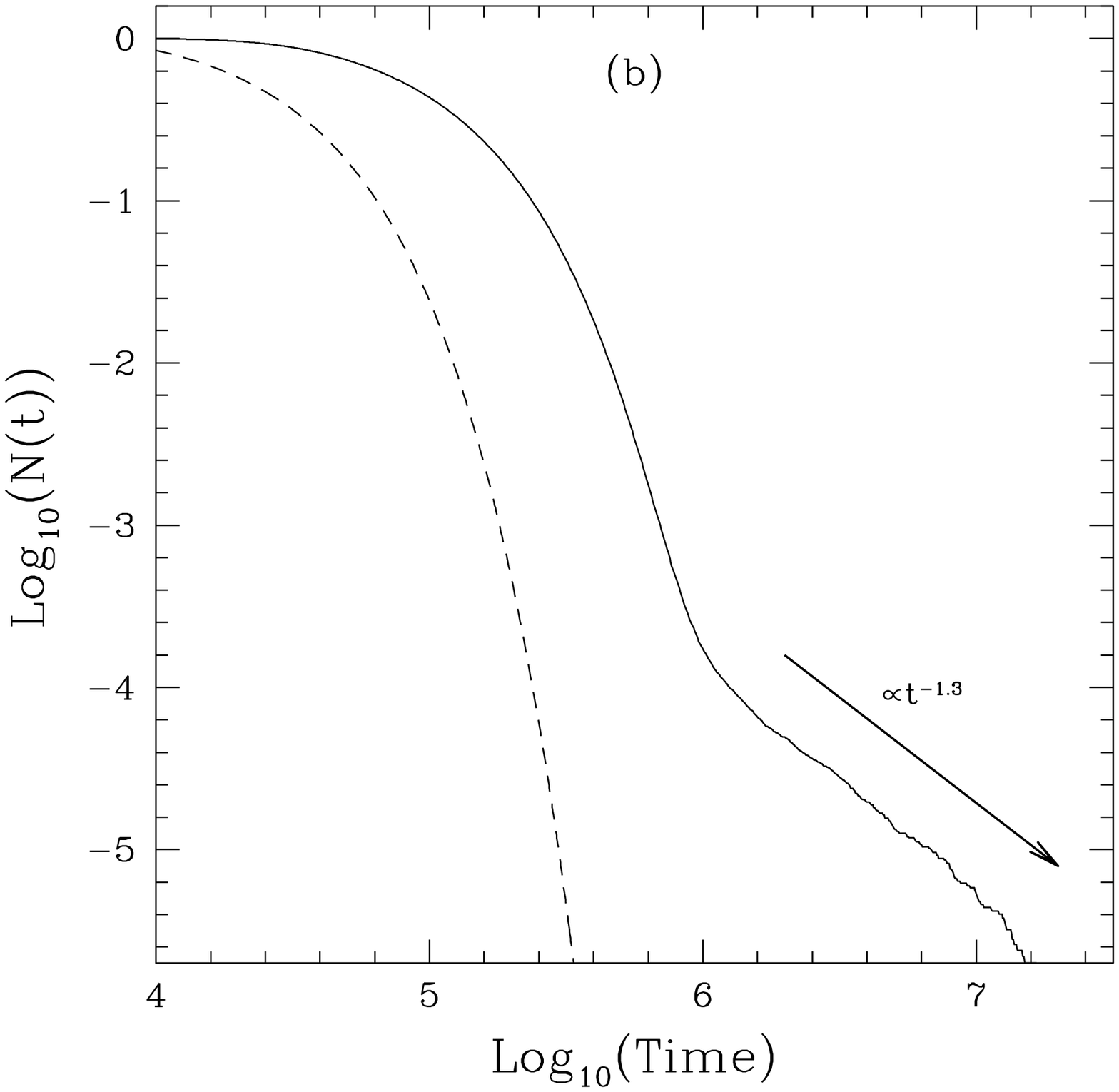}
\caption{The solid lines show the logarithm of the fraction of surviving
comets in the map (\ref{eq:mapping}) inside the energy interval
$-0.59<E/2\pi^2<-0.39$, as a function of time (a) and logarithm of time (b).
The results are based on an initial sample of $5\times10^6$ comets with
initial energy $E/2\pi^2=-0.49$. The depletion function for the diffusion
approximation~(\ref{eq:diffa}) is shown by a dashed line. The
arrow shows the power law $N(t)\propto t^{-1.3}$.} 
\end{figure}

\begin{figure}
\vspace{9cm}
\includegraphics{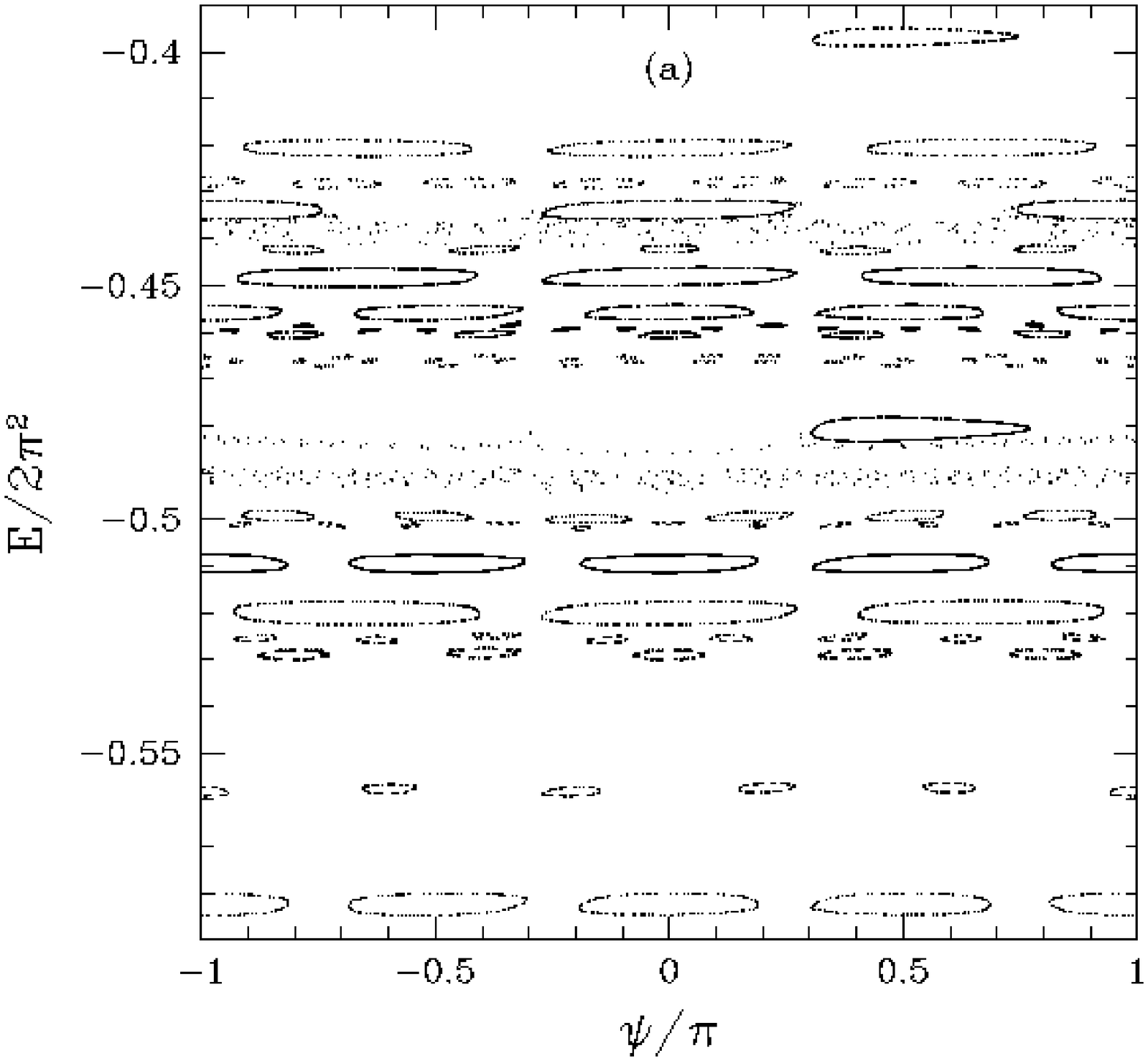}
\includegraphics{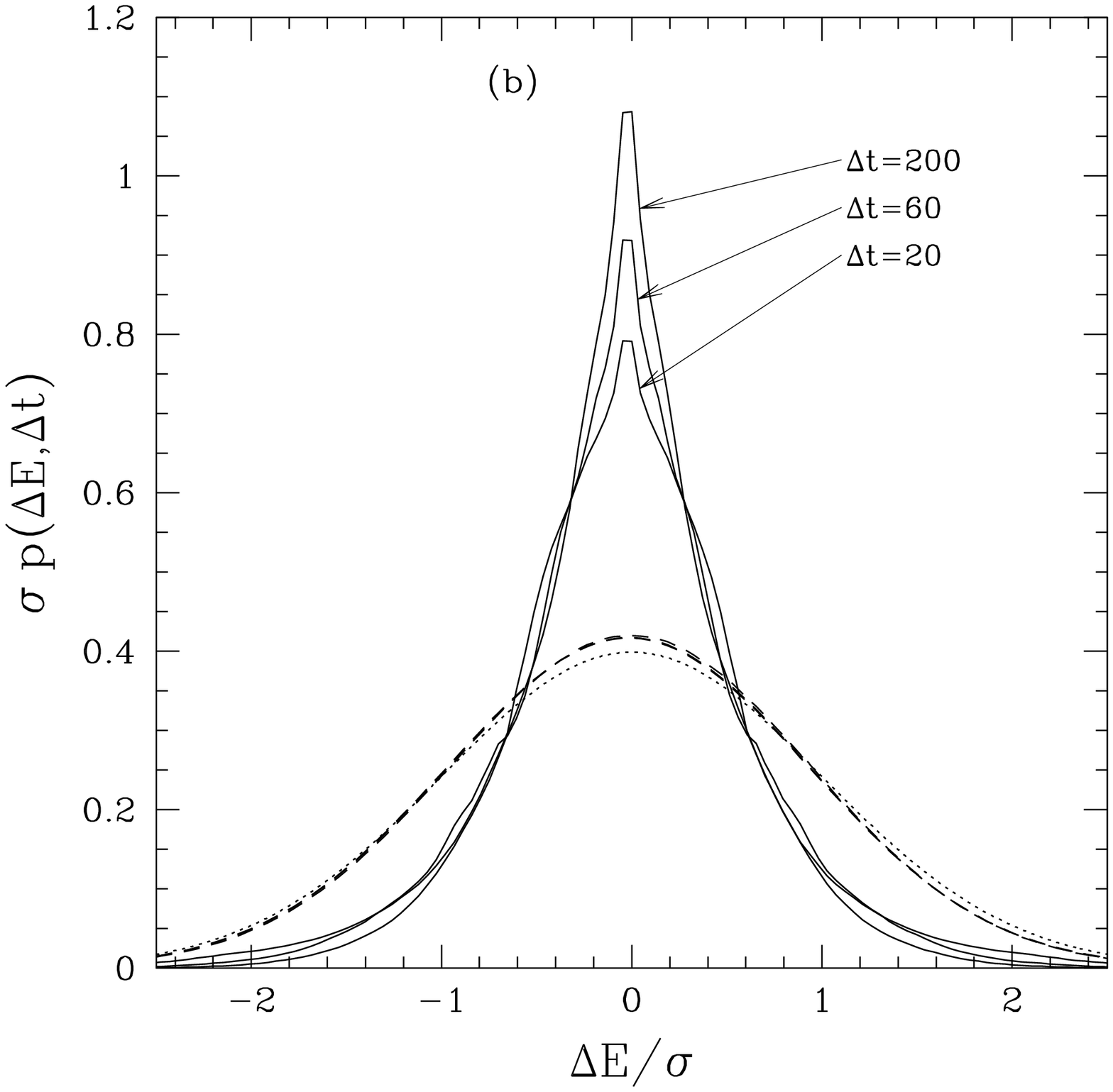}
\caption{(a) The map of the $67$ comets that survived after $5\times 10^6$
Neptune years of the mapping shown in Figure 8. The orbits are stuck to the
edges of the resonant islands. (b) Correlation functions $\sigma\,p(\Delta
E/\sigma)$ for time intervals $\Delta t=20$, 60, 200 (solid lines). The
Gaussian correlation function expected from the diffusion approximation
(eq. \ref{eq:gaussian}) and the correlation function for a Gaussian random walk
with the same boundary conditions as the map are shown by dotted and dashed
lines respectively. }
\end{figure}

\subsection{Other planet masses}

So far we have explored the Keplerian map for planet mass
$m_p=5.24\times10^{-5}$, corresponding to Neptune. Figures 9(a) and (b) show
depletion functions for Saturn ($m_p=2.86\times 10^{-4}$) and Jupiter
($m_p=9.55\times 10^{-4}$). In both cases we calculated time in the
corresponding planet years and comets were lost when they either reached 
semimajor axis $1000$ AU or collided with the Sun ($L\le 0$). The Saturn and
Jupiter maps began with $2\times10^6$ and $10^7$ comets respectively.
The results for all three planet masses are similar: the number of surviving
comets decays approximately exponentially until a time $t_s$, and thereafter
decays as a power law, $N(t)\propto t^{-k}$ with $k\simeq 1.3$. The time $t_s$
appears to be independent of the initial energy: $t_s\simeq 4\times10^7$
planet years for Neptune (from Figure 6[b]), $1.7\times 10^6$ for Saturn, and
$1.5\times 10^5$ for Jupiter. These times can be crudely fit by the formula
$t_s\sim 0.14/m_p^2$. 

Figure~10(a) shows 1000 Jupiter years of the map, for the $47$ comets with
initial energy $E/2\pi^2=-0.5$ that survived for $2\times 10^6$ Jupiter
years. Once again, most of the surviving comets are stuck near resonant
islands.

To compare the depletion functions for Jupiter, Saturn and Neptune we
superimpose these on Figure~10(b) for (approximately) the same initial energy
$E/2\pi^2\simeq -0.24$. Note that the timescale is planet years.

\begin{figure}
\vspace{9cm}
\includegraphics{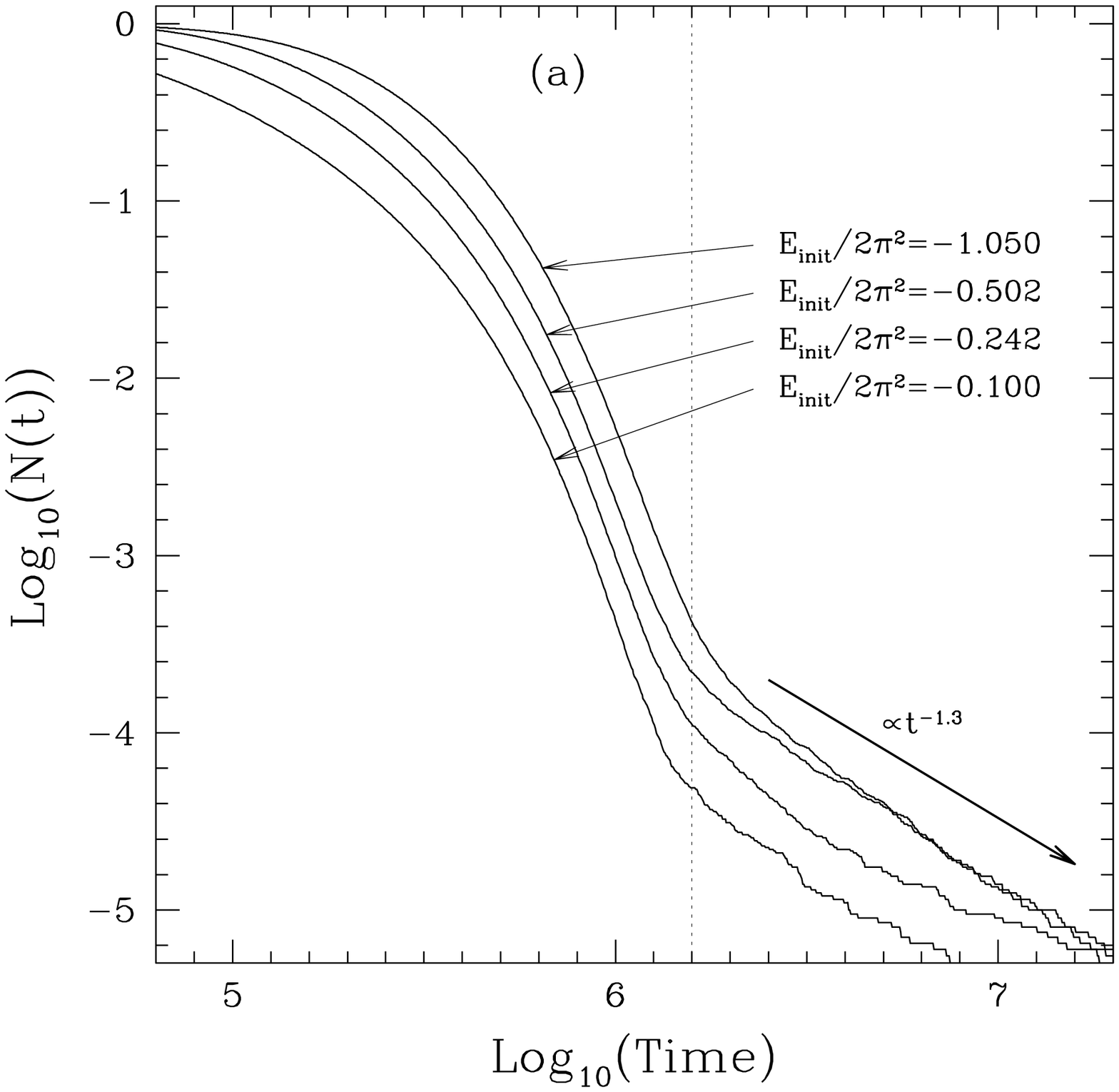}
\includegraphics{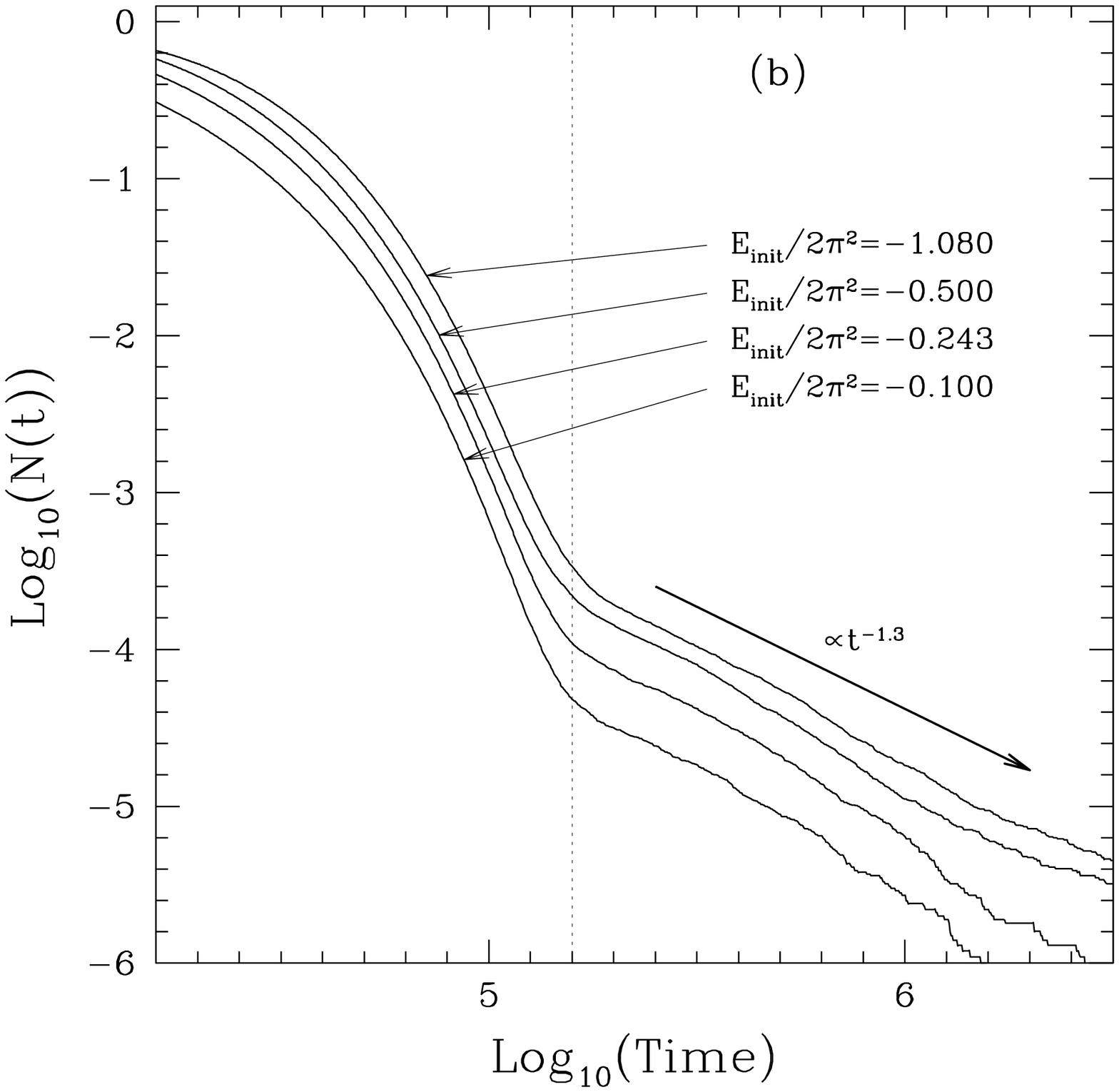}
\caption{ The depletion functions for the Keplerian map with a planet mass
equal to that of Saturn (a) and Jupiter (b). We used different initial
energies for the comets, as indicated in the Figures, but in all cases the
comets were in the stochastic sea. The comets are assumed to escape when they
reach a semimajor axis $1000$ AU or collide with the Sun ($L\le 0$). The
vertical dotted lines represent sticking times defined by equation 
(\ref{eq:stickness_time}).}
\end{figure}

\begin{figure}
\vspace{9cm}
\includegraphics{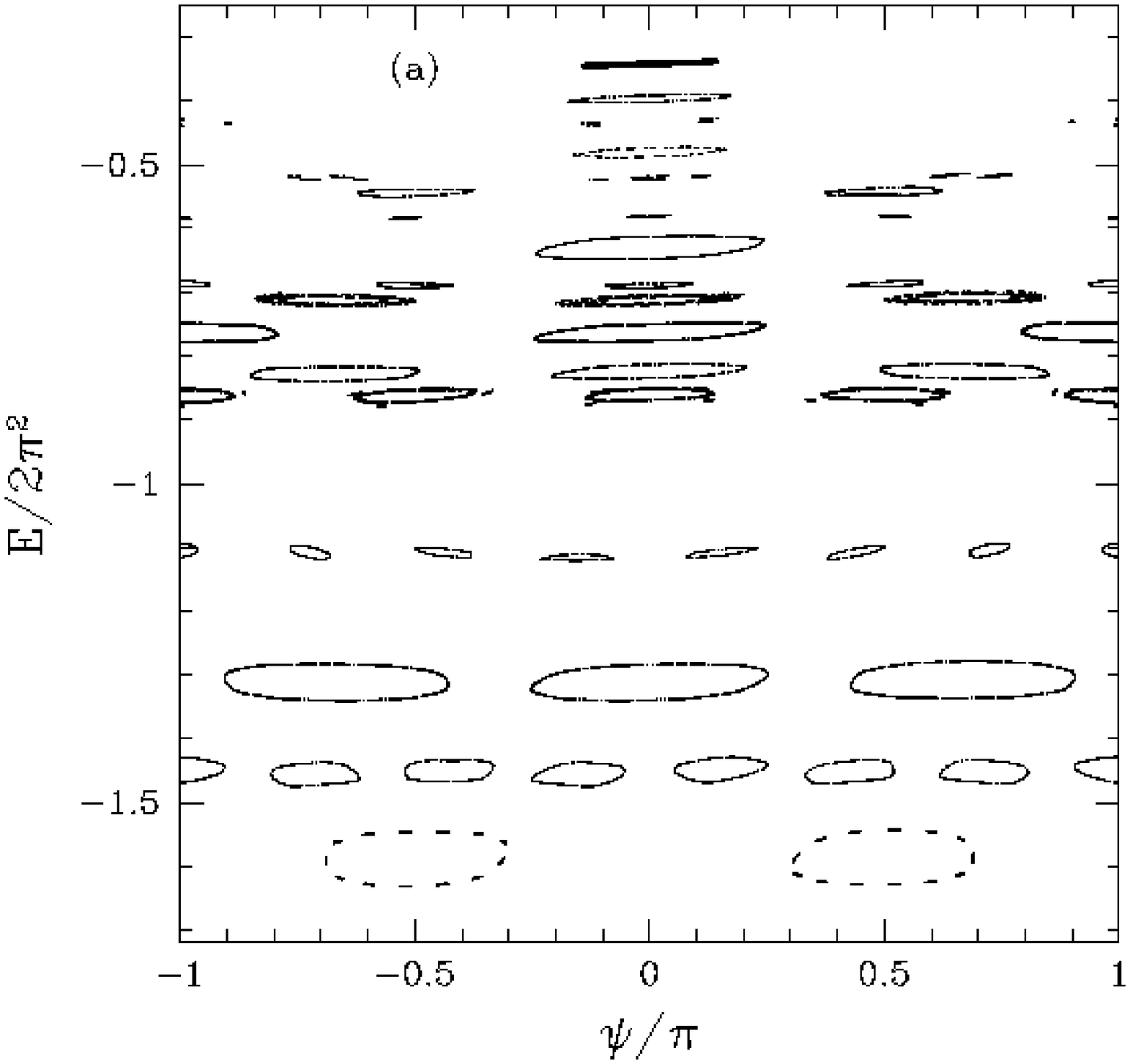}
\includegraphics{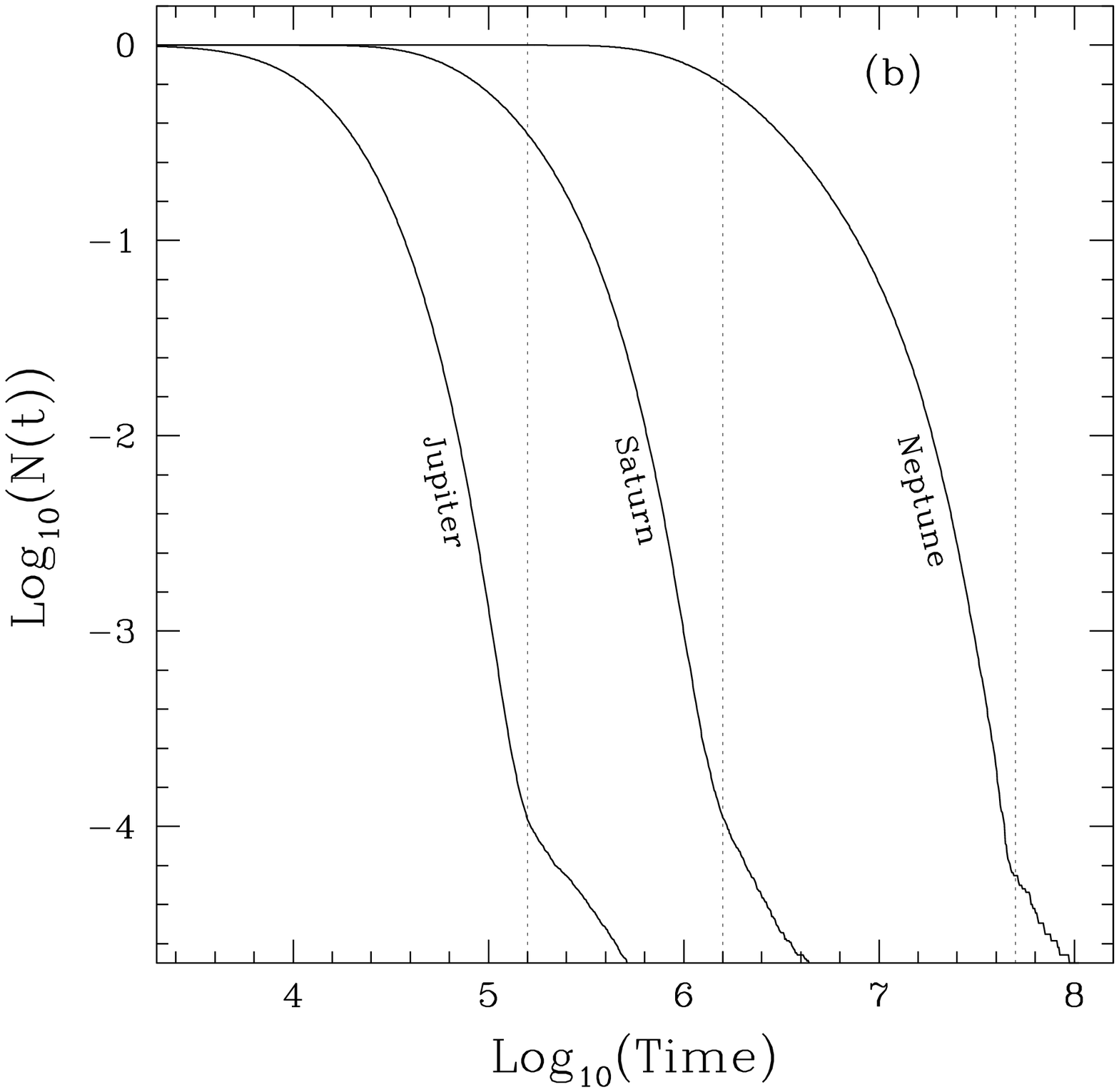}
\caption{ (a) The map of the $47$ comets that survived after $2\times10^6$
planet years of the mapping, with Jupiter's mass and initial comet energy
$E/2\pi^2=-0.5$. The orbits are stuck to the edges of the resonant islands.
(b) A superposition of the depletion functions for mappings with Jupiter,
Saturn and Neptune [from Figures~6(b),~9(a),~9(b)]. The initial comet energy
(in the corresponding planet energy units) was approximately the same in each
case, $E/2\pi^2\simeq -0.24$. The vertical dotted lines represent sticking
times defined by equation (\ref{eq:stickness_time}).}
\end{figure}

\section{Discussion}

\label{sec:disc}

The results of the previous section suggest that the depletion function for
the Keplerian mapping, with absorbing boundaries at $0<E_{max}<E<E_{min}$, 
can be approximately represented in the form 
\begin{eqnarray}
N(t) & \simeq & \exp(-\alpha m_p^2 D_0 t/P),  \qquad t<t_s\\
     & \simeq & \beta {\left(t_s/t\right)}^k, \qquad \qquad t>t_s. 
\label{eq:sstick}
\end{eqnarray}
Here the time $t$ is measured in physical units (e.g. Earth years), $P$ is the
planet's orbital period in Earth years, $\alpha$ and $\beta$ are constants
that are approximately independent of planet mass but may depend
on the initial energy or angular momentum of the comets, $D_0$ is defined by
equation (\ref{eq:diff_coeff}), and $k\simeq 1.3$. This formula only applies
if the escape energy is non-zero. The ``sticking time'' $t_s$ is given by
\be
t_{s}=\frac{\log(1/\beta)}{\alpha D_0}\frac{P}{m_p^2}. 
\label{eq:stickness_time}
\ee 
We show sticking times calculated by equation (\ref{eq:stickness_time}), with
$\alpha\approx 0.04$ and $\beta\approx 10^{-4}$, on Figures~9~and~10(b) as 
vertical dotted lines.  

Our studies of the Keplerian map show that resonance sticking does occur for
highly eccentric planet-crossing orbits. The characteristic sticking time,
after which resonance sticking dominates the depletion function, is $t_s\simeq
2\times 10^6$ (Earth) years for Jupiter, $5\times10^7$ years for Saturn,
$5\times10^9$ years for Uranus, and $7\times10^9$ years for Neptune (for
perihelion distance 0.5 times the planet's semimajor axis). Thus resonance
sticking would be unimportant for a planetary system containing one planet
with the mass and semimajor axis of Uranus or Neptune. However, if the planet
resembled Jupiter or Saturn, resonance sticking would dominate the long-term
behaviour of comets on planet-crossing orbits.

These results invite comparison with long integrations of the orbits of
Neptune-crossing objects. Duncan and Levison (1997) followed 2200 test
particles on Neptune-crossing orbits for $4\times10^9$ years in the
gravitational field of the Sun and the giant planets. They found that on Gyr
timescales the depletion function could be described by a law of the form
(\ref{eq:sstick}) with $k\simeq 1$, normalized so that 1\% of their
particles survive after 4 Gyr.

These results are reminiscent of the results from the map: there is power-law
behaviour at large times, the exponent $k\simeq 1$ is close to the exponent
$1.3$ observed in the map, and the surviving comets at the end of the
integration were mostly in Neptune resonances, just as in the map. However,
our map predicts that sticking is only important for $t>t_s=7\times10^9$
years, and that at $t=t_s$ only $10^{-4}$ of the initial particles survive,
which is smaller by two orders of magnitude. The reason for this discrepancy
between the map and direct integrations is unknown; perhaps the enhanced role
of resonances reflects the extra degrees of freedom in the integration, which
has four planets on eccentric, inclined orbits rather than one on a circular
orbit.

To investigate evolution in more realistic planetary systems, we would like to
generalize the map to include (i) multiple planets; (ii) eccentric planetary
orbits; (iii) dependence of the perturbing Hamiltonian $\kappa$ on energy and
angular momentum, not just the azimuthal angle $\psi$
(cf. eq. \ref{eq:jaccc}). All of these generalizations are straightforward in
principle, although the kick function will depend on more variables and hence
requires a more complicated parametrization. When $\kappa$ depends on $E$ and
$L$, integrating Hamilton's equations across the delta-function at $\ell=0$ is
generally no longer analytic. One possibility is to use a heuristic form for
$\kappa$ chosen so that the integration across $\ell=0$ is analytic. A more
general approach is to work with the mixed-variable generating function that
describes the canonical transformation induced by the perturbing Hamiltonian
across $\ell=0$. Thus if $(L,-E,\varpi,t)$ are the canonical variables just
before perihelion passage at $\ell=0$, primes denote the same variables after
perihelion passage, and $S(L',E',\varpi,t)$ is the generating function, then
\be 
L={\partial S\over\partial\varpi}, \quad E=-{\partial S\over\partial t},
\quad t'=-{\partial S\over\partial E'}, \quad \varpi'={\partial S\over\partial
L}.
\label{eq:canon}
\ee 
It is also straightforward to show that 
\be 
S(L',E',\varpi,t)=L'\varpi -E't-m_p\kappa(L',E',\varpi,t)+\hbox{O}(m_p^2).  
\ee

Equations (\ref{eq:canon}) are implicit in the post-perihelion variables, but
iterative schemes converge rapidly for small planet mass $m_p$. 

\section{Summary}

The Keplerian map offers considerable insight into planet-induced evolution of
highly eccentric orbits, both because of its simplicity and its computational
speed. 

The map reveals that the exponential decay in the number of planet-crossing
comets predicted by the diffusion or random-walk approximation is replaced at
late times by a power-law decay, $N\sim t^{-1.3}$, and that most surviving
comets are ``stuck'' at the edges of resonant islands. Resonance sticking
dominates the evolution of comets under Jupiter's influence after timescales
as short as a few Myr. 

Sticking to Neptune resonances is not strong enough to explain the remarkably
high survival fraction of Neptune-crossing comets (1\% after 4 Gyr) found by
Duncan and Levison (1997), most likely because the simple map used in this
paper does not include eccentric planetary orbits, multiple planets, or
dependence of the planetary perturbations on perihelion distance.

An unresolved issue is how small perturbations and dissipative effects
(passing stars, non-gravi\-ta\-tional forces, etc.) affect the lifetimes of
comets that are stuck to resonances.

We thank Martin Duncan, Matt Holman and Norm Murray for discussions.  This
research was supported in part by NASA grant NAG5-7310.

\end{document}